\documentclass[10pt,a4paper,journal,comsoc, twocolumn]{IEEEtran}

\usepackage[T1]{fontenc}
\usepackage[latin9]{inputenc}
\usepackage{graphicx}
\usepackage{color}
\usepackage{verbatim}
\usepackage{float}
\usepackage{tipa}
\usepackage{amsmath}
\usepackage{amsthm}
\usepackage{amssymb}
\usepackage{graphicx}
\usepackage[unicode=true,
bookmarks=true,bookmarksnumbered=true,bookmarksopen=true,bookmarksopenlevel=1,
breaklinks=false,pdfborder={0 0 0},pdfborderstyle={},backref=false,colorlinks=false]
{hyperref}
\hypersetup{pdftitle={Your Title},
	pdfauthor={Your Name},
	pdfborderstyle={},pdfpagelayout=OneColumn,pdfnewwindow=true,pdfstartview=XYZ,plainpages=false}
\usepackage{epsfig}
\usepackage{amsthm}
\usepackage{algorithm}
\usepackage{algorithmic}

\makeatletter

\pdfpageheight\paperheight
\pdfpagewidth\paperwidth

\newcounter{tempEquationCounter}
\newcounter{thisEquationNumber}
\newenvironment{floatEq}
{\setcounter{thisEquationNumber}{\value{equation}}\addtocounter{equation}{1}
	\begin{figure*}[!t]
		\normalsize\setcounter{tempEquationCounter}{\value{equation}}
		\setcounter{equation}{\value{thisEquationNumber}}
	}
	{\setcounter{equation}{\value{tempEquationCounter}}
		\hrulefill\vspace*{4pt}
	\end{figure*}
}

\newtheorem{theorem}{\bf Theorem}

\newtheorem{definition}{\bf Definition}

\newcommand{\learningratei}{\textcolor{black}{ \iota_{f}^{\text{(1)}}}}
\newcommand{\learningrateii}{\textcolor{black}{ \iota_{f}^{\text{(2)}}}}
\newcommand{\learningrateiii}{\textcolor{black}{ \iota_{f}^{\text{(3)}}}}

\newcommand{\learningrateis}{\textcolor{black}{ \iota^{\text{(1)}2}} } 
\newcommand{\learningrateiis}{\textcolor{black}{ \iota^{\text{(2)}2}}}
\newcommand{\learningrateiiis}{\textcolor{black}{ \iota^{\text{(3)}2}}}
\newcommand{\flowutility}{ \textcolor{black}{\aleph}}
\newcommand{\boltzmanfactor}{\beta} 

\newcommand{\rate}{\textcolor{black}{r} }      
\newcommand{\ratemmWave}{\textcolor{black}{r_f}}       

\newcommand{\arrivalrate}{\textcolor{black} {a}}  
\newcommand{\maximumrate}{\textcolor{black}{ {a}^{\mathrm{max}}}}

\newcommand{\queuebacklog}{ \textcolor{black}{ \boldsymbol{\Xi}} }
\newcommand{\lyapunovdrift}{ \textcolor{black}{ \Delta(\queuebacklog(t))} }

\newcommand{\sidelobegain}{ \Gamma} 

\newcommand{\thermalnoise}{\eta} 
\newcommand{\thermalnoisecov}{\sigma^{2}} 

\newcommand{\radiochain}{R} 

\newcommand{\maximumdelay}{\textcolor{black} {{d}^{\mathrm{max}}}} 
\newcommand{\delayprob}{\epsilon} 
\newcommand{\beamwidth}{\theta} 
\newcommand{\beamsangle}{\omega} 

\newcommand{\controlpara}{\nu} 

\newcounter{myRefCount}

\makeatother
\begin{document}

\title{\Large Joint Path Selection and Rate Allocation Framework \\ for 5G Self-Backhauled mmWave Networks}\vspace{-5em}
\author{\normalsize Trung~Kien~Vu,~\IEEEmembership{Student~Member,~IEEE,} Mehdi~Bennis,~\IEEEmembership{Senior~Member,~IEEE},\\
	   M{\'e}rouane~Debbah,~\IEEEmembership{Fellow,~IEEE,} and~Matti~Latva-aho,~\IEEEmembership{Senior~Member,~IEEE}

\thanks{Manuscript received May 15, 2018; revised Sept. 21, 2018 and Jan. 18, 2019 and Feb. 18, 2019; accepted Feb. 28, 2019 for publication in IEEE Transactions on Wireless Communications. This research has been financially supported by the Academy of Finland 6Genesis Flagship (grant 318927). The Academy of Finland funding via the grant 307492 and the CARMA grants 294128 and 289611, the Nokia Foundation, the Tauno T$\ddot{\text{o}}$nning Foundation, the Tekniikan edistamiss$\ddot{\text{a}}$$\ddot{\text{a}}$ti$\ddot{\text{o}}$ are also acknowledged. }

\thanks{T. K. Vu, M. Bennis, and M. Latva-aho are with the Centre for Wireless Communications, University of Oulu, 90014 Oulu, Finland, (email: \{trungkien.vu, mehdi.bennis, matti.latva-aho\}@oulu.fi).}

\thanks{M. Debbah is with the Large Networks and System Group (LANEAS), CentraleSup\'elec, Universit\'e Paris-Saclay, 91192 Gif-sur-Yvette, France, and also with the Mathematical and Algorithmic Sciences Laboratory, Huawei France R\&D, 92100 Paris, France (e-mail: merouane.debbah@huawei.com).}
\thanks{This paper was presented in part at the IEEE WCNC 2018 conference in Barcelona, Catalonia, Spain, April, 2018~\cite{Vu2018_WCNC}.}}

\maketitle
\vspace{-5em}
\begin{abstract}
 Owing to severe path loss and unreliable transmission over a long distance at higher frequency bands, this paper investigates the problem of path selection and rate allocation for multi-hop self-backhaul millimeter wave (mmWave) networks. Enabling multi-hop mmWave transmissions raises a potential issue of increased latency, and thus, this work aims at addressing the fundamental questions: \textquotedblleft \textit{how to select the best multi-hop paths and how to allocate rates over these paths subject to latency constraints}?\textquotedblright . In this regard, a new system design, which exploits multiple antenna diversity, mmWave bandwidth, and traffic splitting techniques, is proposed to improve the downlink transmission. The studied problem is cast as a network utility maximization, subject to an upper delay bound constraint, network stability, and network dynamics. By leveraging stochastic optimization, the problem is decoupled into: $\left(i\right)$ path selection and $\left(ii\right)$ rate allocation sub-problems, whereby a framework which selects the best paths is proposed using reinforcement learning techniques. Moreover, the rate allocation is a non-convex program, which is converted into a convex one by using the successive convex approximation method. Via mathematical analysis, a comprehensive performance analysis and convergence proof are provided for the proposed solution. Numerical results show that the proposed approach ensures reliable communication with a guaranteed probability of up to $99.9999\%$, and reduces latency by $50.64\%$ and $92.9\%$ as compared to \textit{baseline models}. Furthermore, the results showcase the key trade-off between latency and network arrival rate.
\end{abstract}


\IEEEpeerreviewmaketitle

\section{Introduction}
The fifth generation (5G) wireless systems are expected to reach multiple gigabits per second (Gbps) and to serve a massive number of wireless-connected devices \cite{5GWhat, Quacomm2016}. In this regard, both academia and industry have paid \textcolor{black}{tremendous} attention to the underutilized \textcolor{black}{mmWave} frequency bands ($30-300$ GHz) due to the current scarcity of wireless spectrum \cite{5GWhat,2013millimeter,2017capacity}. Moreover, the \textcolor{black}{above challenges} can be achieved by; $\left(i\right)$ advanced spectral-efficient techniques, e.g., massive multiple-input multiple-output (MIMO) \cite{2016massive}; and $\left(ii\right)$ ultra-dense self-backhauled small cell (SC) deployments~\cite{2017capacity,2015SmallCell,Vu_LB}. Indeed, massive MIMO has been recognized as one of the promising $5$G techniques, which allows to form highly directional beamforming to utilize the mmWave frequency bands and to provide wireless backhaul for SC deployment~\cite{2015SmallCell,Vu_LB}. Ultra dense SC effectively increases network capacity and coverage in which advanced full-duplex (FD) potentially doubles spectral efficiency and reduces latency \cite{2017capacity,2014network, vu2015cooperative}. In addition to the unprecedented growth of data traffic and devices, the issues of low-latency and high-reliability represent other important concerns in 5G networks and beyond \cite{5GWhat,vu2017ultra,2018ultra,2018URC, 2014ultraRC, 2015URLLC, 2018ultra_petar}.

This paper investigates the above $5$G enablers, namely mmWave communication, massive MIMO, and ultra-dense SC deployment, envisaged as the key promoters for providing Gbps data rate, low latency, and highly reliable communication~\cite{5GWhat,2017capacity}. In particular, an in-band access and wireless backhauling are considered to enable the ultra-dense SC deployment~\cite{2015SmallCell,Vu_LB,2008In-band,2015matching} by combining massive MIMO and mmWave to provide Gigabits capacity for both  access and wireless backhaul \cite{2016massive,2013beamforming}.  Owing to the short wavelength, mmWave frequency bands allow for packing a massive number of antennas into highly directional beamforming over a short distance \cite{2013beamforming, 2014channelestimate, 2016hybrid}. Besides that, transmitting over a long distance, mmWave communication requires higher transmit power and is very sensitive to blockage \cite{2013millimeter,Vu_LB,2017capacity}. Hence, instead of using a single hop \cite{Vu_LB,vu2017ultra}, a multi-hop self-backhauling architecture is a promising solution to enable transmissions over long distances in $5$G mmWave networks~\cite{Quacomm2016}, \cite{2009multihop}.  However, using multi-hop transmissions raises the critical issue of increased delay, which has been generally ignored \cite{2009multihop,2012cross_multihop,2016mmW_D2D,2016routing_scheduling,2017multihop,2011PS, 2017gbps}. Unavoidably, ultra-dense SC network is mainly operated based on the multi-hop multi-path transmission fashion~\cite{2017capacity,2014network}, \cite{Quacomm2016}. Hence, there is a need for fast and efficient multi-hop scheduling with respect to traffic dynamics and channel variances in $5$G self-backhauled mmWave networks \cite{Quacomm2016,2015survey}. These previous works focused on addressing one or few issues, have not studied the problem a joint path selection (\textbf{PS}) and rate allocation (\textbf{RA}) in mmWave networks to ensure Gbps data rate and low latency with reliable communications. Thus far, to the best of our knowledge, we are perhaps the first to provide a theoretical and practical framework for addressing all these above concerns.
\vspace{-1em}
\subsection{\textit{Main contributions}}
\textcolor{black}{In this work,  a new system design, which exploits multi-hop transmission, multiple antenna diversity, mmWave bandwidth, and dynamic \textbf{PS} with traffic splitting techniques, is proposed to overcome the severe path loss and mitigate the impact of blockage. The main contributions of the work are listed as follows:}
\begin{itemize}
\item \textcolor{black}{A joint \textbf{PS} and \textbf{RA} optimization for multi-hop multi-path scheduling is formulated, whereby self-backhauled FD SCs act as relay nodes to forward data from the macro BS to the intended UEs. Multi-hop transmission technique enables reliable mmWave communications over a long distance. However, there is a probability that the mmWave signal can be blocked by the human body. Hence, we also introduce the multi-path selection scheme in which the transmitter smartly selects a subset of the best paths among the possible paths.}

\item \textcolor{black}{In the proposed system design, leveraging massive array antenna, hybrid beamforming is adopted to provide Gbps data rate at mmWave bands. In addition, we impose a probabilistic latency bound to ensure URLLC with high data rate. For this purpose, the studied problem is cast as a network utility maximization (NUM), subject to a bounded latency constraint and network stability.}

\item Leveraging stochastic optimization framework \cite{neely2010S}, the studied problem is decoupled into two sub-problems, namely \textbf{PS} and \textbf{RA}. By utilizing the benefits of historical information, a reinforcement learning (RL) is used to build an empirical distribution of the system dynamics to aid in learning the best paths to solve \textbf{PS} \cite{2011learning,bennis2013self}. Therein, the concept of regret strategy is employed, defined as the difference between the average utility when choosing the same paths in previous times, and its average utility obtained by constantly selecting different paths \cite{2011learning, bennis2013self}. The premise is that regret is minimized over time so as to choose the best paths. Second, to solve a non-convex \textbf{RA} sub-problem,  the concept of successive convex approximation (SCA) method is applied due to its low complexity and fast convergence \cite{beck2010seq, tran2012}.

\item The proposed approach answers the following fundamental questions: $(i)$ \textit{over which paths should the traffic flow be forwarded}? and $(ii)$ \textit{what is the data rate per flow/sub-flow?,} while ensuring a probabilistic delay constraint, and network stability. By using a mathematical analysis, a comprehensive performance of our proposed stochastic optimization framework is scrutinized. It is shown that there exists an $[\mathcal{O}(1/\nu),\mathcal{O}(\nu)]$ utility-queue backlog trade-off, which leads to an utility-delay balancing~\cite{neely2010S}, where $\nu$ is a control parameter. In addition, a convergence analysis of both two sub-problems is studied. Finally, the performance of the proposed solution is validated by extensive set of simulations.
\end{itemize}
\vspace{-1.5em}
\subsection{\textit{Related work}}
A tractable rate model was proposed to characterize the rate distribution in self-backhauled  mmWave networks \cite{2015tractable}. Few efforts have been made to study the mmWave network operation regime, noise-limited or interference limited, depending on the density of interferers, transmission strategies, or channel propagation models \cite{2015transit, 2016regimes, 2017INR}. A large body of research work has attempted to study the joint \textbf{RA}, congestion control, routing, and scheduling for multi-hop wireless networks, incorporating the proportional delay based on the sum of queue backlogs \cite{2012cross_multihop}, applying the concept of back-pressure algorithm \cite{2011novel_back-pressure,2014Uti_delay}, exploiting the potential of multiple gateways \cite{2016routing_scheduling}.

The authors in \cite{2015analysis} considered a problem of joint scheduling and congestion control in a multi-hop mmWave network using a NUM framework in which the proposed solution is verified under three interference models, namely graph-based actual interference, free-interference (IF), and the worse-case interference. \cite{2015analysis} also showed that the IF model provides very tight upper bound for a realistic system evaluation in mmWave cellular networks as long as the optimal throughput can be guaranteed. However, \cite{2015analysis, vu2014mobility} was concerned only with the network capacity maximization and single path streaming, a tight latency and reliable constraint should be investigated together with dynamic path diversity. Moreover, the authors in \cite{2010delay} designed a multi-hop wireless backhaul scheme with delay guarantee in which a link activation scheme was proposed to avoid interference and minimize the latency. A rate allocation problem to minimize the application layer video/end-to-end distortion subject to quality of service constraints (delay, backhaul) was considered in \cite{2007media, 2009path} for multi-path networks. However, other important aspects in $5$G networks such as low-latency and high-reliability are generally ignored when maximizing the network performance (capacity, energy efficiency and spectral efficiency) \textcolor{black}{\cite{2017gbps, 2015tractable, nguyen2017distributed, ngo2013energy}}.

A recent work in \cite{2017multihop} has studied the multi-hop relaying transmission challenges for mmWave systems, aiming at maximizing overall network throughput, and taking account of traffic dynamics and link qualities. In our work, we also study the NUM optimization problem, while considering channel variations and network dynamics. Another recent work in \cite{2017traffic_multihop} has addressed the problem of traffic allocation for multi-hop scheduling in mmWave networks to minimize the end-to-end latency, in which the minimum latency is derived based on the channel capacity to determine the portions of traffic over channels such that all traffic fractions arrive simultaneously at the destination. \textcolor{black}{In addition, the problem of \textbf{PS} and multi-path congestion control for data transfers was studied in \cite{2011PS} in which the aggregate utility is increased as more paths are provided. One important suggestion is to re-select randomly from the set of paths and shift between paths with higher payoff. However, splitting data into too many paths leads to increased signaling overhead and causes traffic congestion. While interesting, the preceding works do not address the problem of high-data rate, low-latency and reliability communication in multi-path mmWave networks. In this respect, our proposed solution is to select the best paths to maximize the network throughput, subject to a delay bound violation constraint with a tolerable probability (reliability).  Our previous work \cite{vu2017ultra} studied URLLC-centric mmWave networks for single hop transmission, and \cite{Vu_LB} proposed an integrated access and backhaul architecture for two-hop relay without considering the delay-sensitive constraint. Hence, in this work the authors extend to the multi-hop wireless backhaul scenario, and study a joint \textbf{PS} and \textbf{RA} problem focusing on  URLLC.} Via mathematical analyses and extensive simulations, the authors provide insights into the performance analysis of our proposed algorithm and the convergence characteristics of the learning algorithm and the SOCP based iterative method.

The rest of the paper is organized as follows\footnote{${\it \textit{Notations}}$: Throughout the paper, the lowercase letters, boldface lowercase letters, (boldface) uppercase letters and italic boldface uppercase letters are used to represent scalars, vectors, matrices, and sets, respectively. For a matrix $\mathbf{X}$, we use $\mathbf{X}^{\text{T}}$, $\mathbf{X}^{\dag}$ and $\text{Rank}(\mathbf{X})$ to denote its transpose, Hermitian and rank, respectively. $\mathbb{E}[\cdot]$ denotes the expectation operator, $\mathbb{I}_{\{z\}}$ is the indicator function for logic $z$, and $[x]^{+}\triangleq\max\{x,0\}$. The cardinality of a set $\mathcal{S}$, is denoted by $|\mathcal{S}|$. We denote the previous hop and the next hop from node $i$ as $i^{({\rm I)}}$ and $i^{({\rm o)}}$, respectively. $\text{Pr}(\cdot)$ denotes the probability operator.}. Section~\ref{SystemModel} describes the system model and Section~\ref{Pro-Form} provides the problem formulation for a joint  \textbf{PS} and \textbf{RA} optimization. Section~\ref{Design} introduces a stochastic optimization framework to decouple our studied problem, whereby two practical solutions are proposed. A mathematical analysis of the proposed framework is discussed in Section \ref{performance_analysis}. Section~\ref{Evaluation} provides extensive numerical results to compare again other baselines. Conclusions are drawn in Section~\ref{Conclusion}.
\section{System Model}
\label{SystemModel}
\vspace{-0.5em}
\subsection{Network Model}
\label{NetworkModel}
\begin{figure}
\centering\includegraphics[width=1\columnwidth]{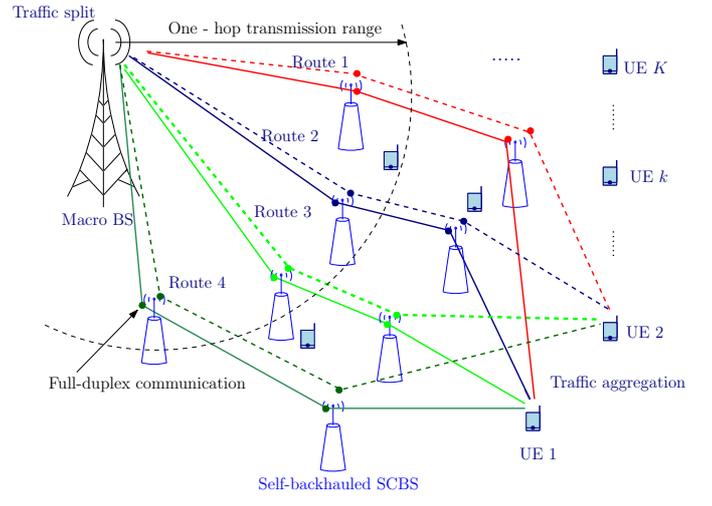}
\caption{Illustration of 5G multi-hop self-backhauled mmWave networks.}
\label{Example-1}
\end{figure}

Let us consider a downlink (DL) transmission of a multi-hop heterogeneous cellular network (HCN) which consists of a macro base station (MBS), a set of $B$ self-backhauled small cell base stations (SCBSs), and a set ${\cal K}$ of $K$ user equipments (UEs) as shown in Fig \ref{Example-1}. Let ${\cal B}=\{0,1,\cdots,B\}$ denote the set of all BSs in which index ${\rm 0}$ refers to the MBS. The in-band wireless backhaul is used to provide backhaul among BSs \cite{2008In-band,Vu2016}. A full-duplex (FD) transmission protocol is assumed at SCBS with perfect self-interference cancellation (SIC) capabilities \cite{2015full, Vu2016, 2017Elbamby}. Each BS $b$ is equipped with $N_{b}$ transmitting antennas and $\radiochain_{b}$ radio frequency (RF) chains, \textcolor{black}{such that $1 \leq \radiochain_{b} \leq N_{b},  \forall b \in \cal{B} $} \cite{2014channelestimate, 2016hybrid, 2015limited}. Similarly, each UE $k$ is equipped with $N_{k}$ transmitting antennas and $\radiochain_{k}$ RF chains, \textcolor{black}{such that $1 \leq \radiochain_{k} \leq N_{k}$, $\radiochain_{k} \leq \radiochain_{b}$, and $N_{k} \ll N_{b}$, $\forall k \in \cal{K}$, $\forall b \in \cal{B}$}. The network topology is modeled as a directed graph $\mathcal{G}=(\mathcal{N},\,\mathcal{L})$, where ${\cal N}=\mathcal{B}\,\cup\mathcal{K}$ represents the set of nodes including BSs and UEs. ${\cal L}=\{(i,j)|i\in\mathcal{B},j\in\mathcal{N}\}$ denotes the set of all directional edges $(i,j)$ in which nodes $i$ and $j$ are the transmitter and the receiver, respectively.

\begin{table*}
\centering\caption{Notations for system model.}
\begin{tabular}{|c|c|}
\hline
Notations & Descriptions\tabularnewline
\hline
\hline
${\cal B}$,${\cal \,{\cal K}}$ & Sets of $\left(B+1\right)$ base stations, $K$ user equipments\tabularnewline
\hline
${\cal {\cal {\cal N}=\mathcal{B}\,\cup\mathcal{K}}}$ & Set of nodes including BSs and UEs\tabularnewline
\hline
${\cal L}$ & Set of all directional edges $(i,j)|i\in\mathcal{B},j\in\mathcal{N}$\tabularnewline
\hline
${\cal F}$ & Set of $F$ flows\tabularnewline
\hline
$\mathcal{Z}_{f}$ & Set of $Z_{f}$ disjoint paths observed by flow $f$\tabularnewline
\hline
$\mathcal{Z}_{f}^{m}$ & Disjoint path state/table $m$ observed by flow $f$\tabularnewline
\hline
$\mathcal{N}_{i}^{({\rm o)}}$  & Set of the next hops from node $i$\tabularnewline
\hline
$i_{f}^{({\rm I)}}$ & Previous hop of flow $f$ to BS $i$ \tabularnewline
\hline
$i_{f}^{({\rm o)}}$ & Next hop of flow $f$ from BS $i$ \tabularnewline
\hline
$p_{(i,j)}^{f}$ & Transmit power of node $i$ to node $j$ for flow $f$\tabularnewline
\hline
$z_{f}^{m}=1$ & Path $m$ is used to send data for flow $f$\tabularnewline
\hline
$\pi_{f}^{m}$ & Probability of choosing path $m$ for flow $f$\tabularnewline
\hline
\end{tabular}\label{notations1}
\end{table*}
We consider a queuing network operating in discrete time $t\in\mathbb{Z}^{+}$. There are $F$ independent data flows at the MBS. Each data traffic is destined for only one UE, whereas one UE can receive up to $\radiochain_k$ multiple data streams, i.e., $F\geq K$. The number of total data streams at the MBS is no greater than the number of RF chains, \textcolor{black}{such that $F \times \radiochain_k \leq \radiochain_b$, $\forall k \in \cal{K}$, $\forall b \in \cal{B}$} \cite{2016hybrid, 2015limited}. Hereafter, we refer to data traffic as data flow. We use $\mathcal{F}$ to represent the set of $F$ data flows/sub-flows. The MBS can split each flow $f\in{\cal F}$ into multiple sub-flows which are delivered via disjoint paths and aggregated at UEs~\cite{2016traffic, 2016multiC}.

We assume that there exists $Z_{f}$ number of disjoint paths from the MBS to the UE for flow $f$. For any disjoint path $m\in\left\{ 1,\cdots,Z_{f}\right\} $, we denote $\mathcal{Z}_{f}^{m}$ as the path state, which contains all path information such as topology and queue states for every hop. Let $\mathcal{Z}_{f}=\{\mathcal{Z}_{f}^{1},\cdots,\mathcal{Z}_{f}^{m},\cdots,\mathcal{Z}_{f}^{Z_{f}}\}$ denote the path states/tables observed by flow $f$. We use the flow-split indicator vector $\mathbf{z}_{f}=\left(z_{f}^{1},\cdots,\,z_{f}^{Z_{f}}\right)$ to denote how the MBS splits flow $f$, where $z_{f}^{m}=1$ means path $m$ is used to send data for flow $f$; otherwise, $z_{f}^{m}=0$. Let $\mathcal{N}_{i}^{({\rm o)}}$ denote the set of next hops from node $i$ via a directional edge. We denote the next hop and the previous hop of flow $f\,$ from and to BS $i$ as $i_{f}^{({\rm o)}}$and $i_{f}^{({\rm I)}}$, respectively. Table \ref{notations1} shows the notations, used throughout this paper.

\vspace{-1em}
\subsection{mmWave MIMO Channel Model}
\label{channel_model}
\textcolor{black}{Due to limited spatial scattering in mmWave MIMO propagation \cite{2013millimeter, 2015limited}, we assume that there are $L_{(i,j)}$ clusters between transmitter $i$ and receiver $j$, such that $L_{(i,j)} \ll \min (N_i,N_j)$. The channel matrix ${{\bf H}}_{(i,j)}$ of link $(i,j)$ can be modelled as \cite{2015limited, 2015beamsearch, 2018editor}
\begin{equation}\label{mmWavechannel0}
{{\bf H}}_{(i,j)}=\sqrt{ \frac{N_{i} \times N_j }{L_{(i,j)}} } \sum_{l=1}^{L_{(i,j)}} {h}_{(i,j)}(l)  {\bf A}_{j}(\alpha_{j, l}) {\bf A}_{i}^{\dagger}(\alpha_{i, l}),
\end{equation}
where ${h}_{(i,j)}(l)$ denotes the small-scale fading coefficient of the cluster $l^{\text{th}}$. $\alpha_{j, l}$ and $\alpha_{i, l}$ denote the azimuth angles of arrival and departure, respectively. Here, $ {\bf A}_{i}(\alpha_{i, l})$ and $ {\bf A}_{j}(\alpha_{i, l})$ represent the transmitter and receiver response vectors, respectively (Please refer \cite{2015beamsearch, 2018editor} for more details).  We denote ${\bf H}= \big \{ {\bf H}_{(i,j)}|(i,j)\in\mathcal{L} \big \}$ as the network channel matrix.}

\vspace{-1em}
\subsection{Rate Formulation}
\label{rate_formulation}
We denote $p_{(i,j)}^{f}$ as the transmit power of node $i$ assigned to node $j$ for flow $f$, such that $\sum_{f\in F}\sum_{j\in{\cal N}_{i}^{{\rm (o)}}}p_{(i,j)}^{f}\leq P_{i}^{{\rm max}},$ where $P_{i}^{{\rm max}}$ is the maximum transmit power of node $i$. We have the following power constraint
\vspace{-0.5em}
\begin{align}
\!\mathcal{P} & =\bigg\{ p_{(i,j)}^{f}\geq0,i,j\in\mathcal{N},\Big|\sum_{f\in{\cal F}}\sum_{j\in{\cal N}_{i}^{{\rm (o)}}}p_{(i,j)}^{f}\leq P_{i}^{{\rm max}}\bigg\}.\label{eq:powerconstraint-0}
\end{align}
Vector $\mathbf{p}=(p_{(i,j)}^{f}|\forall i,j\in{\cal N},\forall f\in{\cal F})$ denotes the transmit power over all flows.

Based on the hybrid beamforming and combining model \cite{2016hybrid, 2015limited}, with ${\bf c}_{(i,j)} \in \mathbb{C}^{N_{j} \times 1}$ as the RF combining and baseband equalizer and ${\bf v}_{(i,j)} \in \mathbb{C}^{N_{i} \times 1}$ as hybrid analog/digital precoding,  the Ergodic achievable rate\footnote{Note that we omit the beam search/tracking time, since it can be done fast and is negligible compared to the transmission
time \cite{2017tracking}.  \textcolor{black}{Due to the disjoint path assumption and directional beamforming, the interference associated to transmissions from transmitter $i$ to other receivers $j^{\prime}$, received at $j$, is assumed to be negligible or can be mitigated by designing the two-layer precoder at the transmitter $i$ \cite{Vu_LB,Liu2014}. For the sake of simplification, the impact of this interference is left for future work. }} $\ddot{\rate}^{(i,j)}$ at the receiver $j$ from the transmitter $i$ can be calculated as \eqref{rate_calculate0}.
\begin{floatEq}
\begin{equation}\label{rate_calculate0}
 \ddot{\rate}^{(i,j)}  =\mathbb{E}_{{\bf H},\,{\bf p}}\left[\text{w}\log\left(1+\frac{p_{(i,j)}\vert {\bf c}_{(i,j)}^{\dagger} {\bf H}^{\text{T}}_{(i,j)}{\bf v}_{(i,j)}\vert^{2}}{\sum_{i^{\prime}\neq i} \textcolor{black}{ \sum_{j^{\prime} \in \mathcal{N}_{i^{\prime}}^{({\rm o)}} }} p_{(i^{\prime},j^{\prime})} \vert {\bf c}_{(i,j)}^{\dagger} {\bf H}^{\text{T}}_{(i^{\prime},j)}{\bf v}_{(i^{\prime},j)}\vert^{2}+ \thermalnoisecov_{j} \Vert{\bf c}_{(i,j)} \Vert^{2}}\right)\right].
\end{equation}
\end{floatEq}
Here $p_{(i,j)}$ is the transmit power from the transmitter $i$  assigned to the receiver $j$, and the
thermal noise of receiver $j$ is $\thermalnoise_{j}\thicksim \mathcal{CN}\left(0, \thermalnoisecov_{j}\right)$.
In addition, $\text{w}$ denotes the system bandwidth of the mmWave frequency band. For a given channel state and transmit power, the data rate in the edge $(i,j)$ over flow $f$ can be posed as a function of channel state and transmit power, i.e., $\ratemmWave^{(i,j)}\left({\bf H},{\bf \,p}\right)$, such that $\sum_{f\in{\cal F}}\ratemmWave^{\left(i,j\right)}= \rate^{\left(i,j\right)}$. We denote $\mathbf{r}=(\ratemmWave^{\left(i,j\right)}|\forall i,j\in{\cal N},\forall f\in{\cal F})$ as a vector of data rates over all flows.

As studied in \cite{2017hybrid}, the previous works on mmWave hybrid beamforming are mainly focused on the physical layer or signal processing aspects \cite{2014channelestimate, 2016hybrid, 2015limited, 2014sparse}. The authors in \cite{2017hybrid} developed an accurate analytical model that captures the essence of mmWave hybrid beamforming, while tractable enough to analyze the throughput-delay performance. In our work, we adopt the model in  \cite{2017hybrid} to formulate the network utility maximization subject to the congestion control and network stability. In particular, let $g_{(i,j)}^{(t)}$ and $g_{(i,j)}^{(r)}$ denote the transmitter and receiver analog beamforming gain at the transmitter $i$ and the receiver $j$, respectively. In addition, we use $\beamsangle_{(i,j)}^{(t)}$ and $\beamsangle_{(i,j)}^{(r)}$ to represent the angles deviating from the strongest path between the transmitter $i$ and the receiver $j$. Also, let $\beamwidth_{(i,j)}^{(t)}$ and $\beamwidth_{(i,j)}^{(r)}$ denote the beamwidth at the transmitter $i$ and the receiver $j$, respectively. We adopt the widely used antenna radiation pattern  \cite{2015beamsearch, 2017hybrid, 2014beamwidth, perfecto2017millimeter} to determine the beamforming gain as
\vspace{-0.5em}
\begin{align*}
g_{(i,j)}\left(\beamsangle_{(i,j)},\beamwidth_{(i,j)}\right) & =\begin{cases}
\frac{2\pi-\left(2\pi-\beamwidth_{(i,j)}\right)  \sidelobegain }{\beamwidth_{(i,j)}}, & \text{if}\:|\beamsangle_{(i,j)}|\leq\frac{\beamwidth_{(i,j)}}{2},\\
\sidelobegain, & \text{otherwise,}
\end{cases}
\end{align*}
where $0< \sidelobegain \ll 1$ is the side lobe gain. After the beam alignment is done, the receiver sends the pilot sequences to the transmitter. The transmitter estimates the channel and precodes signals, throughout this paper, the effective data rate of link $(i,j)$ $\rate^{\left(i,j\right)}$ is calculated as \eqref{rate_calculate1} in which $g_{(i,j)}^{(s)}$ denotes the spatial channel gain of link $(i,j)$ \cite{2015beamsearch, 2018editor, 2014beamwidth}. Note that after the beam-searching and alignment are done \cite{2015beamsearch,2014beamwidth,2014ieee80211ad, 2011ieee80215} the receiver broadcasts pilot sequences to the transmitters, each transmitter estimates the channel to the corresponding receiver and precodes transmit signal in the DL. With multiple $N_j$ antennas and $R_j$ RF chains, each receiver is capable of receiving multiple data streams from different transmitters using either the main beam or the side lope beam. We assume that the traffic split and aggregation are done ideally, the multiple data streams can be transmitted via different paths.
\begin{floatEq}
\begin{equation}\label{rate_calculate1}
\rate^{\left(i,j\right)}  =\mathbb{E}_{{\bf H},\,{\bf p}}\left[\text{w}\log\left(1+\frac{p_{(i,j)} g_{(i,j)}^{(t)} g_{(i,j)}^{(s)} g_{(i,j)}^{(r)}}{\sum_{i^{\prime}\neq i} \textcolor{black}{ \sum_{j^{\prime} \in \mathcal{N}_{i^{\prime}}^{({\rm o)}} }} p_{(i^{\prime},j^{\prime})} g_{(i^{\prime},j)}^{(t)} g_{(i^{\prime},j)}^{(s)} g_{(i^{\prime},j)}^{(r)} + \thermalnoisecov_{j} } \right)\right].
\end{equation}
\end{floatEq}
\vspace{-1em}
\subsection{Network Queues}

Let $Q_{f}^{i}(t)$ denote the queue length at a BS $i$ at time slot $t$ for flow $f$. The queue length evolution at the MBS $i=0$ is
\vspace{-0.5em}
\begin{equation}
Q_{f}^{i}(t+1)=\Big[Q_{f}^{i}(t)-\sum\limits _{m=1,i_{f}^{({\rm o)}}\in\mathcal{Z}_{f}^{m}}^{Z_{f}}\ratemmWave^{(i,i_{f}^{({\rm o)}})}(t),\:0\Big]^{+}+\arrivalrate_{f}(t),\label{eq:queue1}
\end{equation}
where $\arrivalrate_{f}(t)$ is the data arrival at the MBS during slot $t$, which is i.i.d. over time with a mean value $\bar{\arrivalrate}_{f}$ and is bounded by $\arrivalrate_{f}(t) \leq \maximumrate_f < \infty$. Due to the disjoint paths, for each flow $f\,$ the incoming rate from the previous hop $i_{f}^{({\rm I)}}$ at the SCBS $i$ is either from another SCBS or the MBS, and thus, the queue evolution at the SCBS $i=\left\{ 1,\cdots,\,B\right\} $ is given by
\begin{equation}
Q_{f}^{i}(t+1)=\Big[Q_{f}^{i}(t)-\ratemmWave^{(i,i_{f}^{({\rm o)}})}(t),\:0\Big]^{+}+\ratemmWave^{(i_{f}^{({\rm I)}},i)}(t).\label{eq:queue2}
\end{equation}

\begin{definition}
For any vector $\mathbf{x}\left(t\right)=\left(x_{1}\left(t\right),...,x_{\emph{K}}\left(t\right)\right)$,
let $\bar{\mathbf{x}}=\left(\bar{x}_{1},\cdots,\bar{x}_{\emph{K}}\right)$
denote the time average expectation of $\mathbf{x}\left(t\right)$,
where ${\textstyle \bar{\mathbf{x}}\triangleq\lim_{t\to\infty}\frac{1}{t}\sum_{\tau=0}^{t-1}\mathbb{E}\left[\mathbf{x}\left(\tau\right)\right]}$.
\end{definition}
\begin{definition}
For any discrete queue $Q\left(t\right)$ over time slots $t\in\left\{ 0,1,\ldots\right\} $
and $Q\left(t\right)\in R_{+}$,
\end{definition}
\begin{itemize}
\item $Q\left(t\right)$ is strongly stable if ${\textstyle \lim_{t\to\infty}\sup\frac{1}{t}\sum_{\tau=0}^{t-1}\mathbb{E}\left[|Q\left(\tau\right)|\right]<\infty}$.
\item $Q\left(t\right)$ is mean rate stable if ${\textstyle \lim_{t\to\infty}\frac{\mathbb{E}\left[|Q\left(t\right)|\right]}{t}=0}$.
\end{itemize}
A queuing network is stable if each queue is stable.

\section{Problem Formulation}
\label{Pro-Form}
Assume that the MBS determines which paths to split data flow $f$ with a given probability distribution, i.e., $\boldsymbol{\pi}_{f}=\big(\pi_{f}^{1},\cdots,\pi_{f}^{Z_{f}}\big)$, where for each $m\in{\cal Z}_{f}$ we have $\pi_{f}^{m}=\text{Pr}\left(z_{f}=z_{f}^{m}\right)$.
Here, $\boldsymbol{\pi}_{f}$ is the probability mass function (PMF) of the flow-split vector, i.e., $\sum_{m=1}^{Z_{f}}\text{Pr}\left(z_{f}^{m}\right)=1$.
We denote $\boldsymbol{\pi}=\left\{ \boldsymbol{\pi}_{1},\cdots,\boldsymbol{\pi}_{f},\cdots,\boldsymbol{\pi}_{F}\right\} \in\Pi$ as the global probability distribution of all flow-split vectors in which $\Pi$ is the set of all possible global PMFs. Let $\bar{x}_{f}$ denote the achievable average rate of flow $f\,$ such that
\[
\bar{x}_{f}\triangleq\lim\limits _{t\to\infty}\frac{1}{t}\sum_{\tau=0}^{t-1}x_{f}\left(\tau\right),
\]
\text{and} \,
\[
x_{f}\left(\tau\right)=\sum_{m=1,i_{f}^{({\rm o)}}\in\mathcal{Z}_{f}^{m}}^{Z_{f}}\mathbb{E}_{{\bf H},{\bf p}}\big[\pi_{f}^{m}\ratemmWave^{(i,i_{f}^{({\rm o)}})}(\tau)\big]\Big|_{i=0}.
\]
We assume that the achievable rate is bounded, i.e.,
\begin{align}
0 & \leq x_{f}(t)\leq \maximumrate_{f},\label{ratebounds}
\end{align}
where $\maximumrate_f$ is the maximum achievable rate of flow $f$ at every time $t$. Vector $\bar{{\bf x}}=\left(\bar{x}_{1},\cdots,\bar{x}_{F}\right)$
denotes the time average of rates over all flows. Let $\mathcal{R}$
denote the rate region, which is defined as the convex hull of the
average rates, i.e., $\bar{{\bf x}}\in\mathcal{R}$.

We define $U_{0}$ as the network utility function, i.e., $U_{0}\left(\bar{{\bf x}}\right)=\sum_{f\in{\cal F}}U\left(\bar{x}_{f}\right)$
\cite{2011PS,Vu_LB}. Here, $U(\cdot)$ is assumed to
be a twice differentiable, concave, and increasing $L\text{-Lipschitz}$
function for all $\bar{{\bf x}}\geq0$. According to Little's law
\cite{2008little}, the average queuing delay is defined as the ratio
of the queue length to the average arrival rate. By taking account
of the probabilistic delay constraints for each flow/subflow, the
network utility maximization (NUM) is formulated as follows:
\vspace{-0.5em}
\begin{subequations}\label{OP1}
\begin{align}
\hspace{-1em}\mbox{OP:}\:\max_{\boldsymbol{\pi},\mathbf{\,x},\,{\bf p}} & \quad U_{0}(\bar{\mathbf{x}})\label{eq:obj1}\\
\text{subject to}
& \text{\quad Pr \ensuremath {\Big(\frac{\text{\ensuremath{Q_{f}^{i}(t)}}}{\bar{\arrivalrate}_{f}} \geq \maximumdelay \Big) \leq \delayprob}},\forall t,f\in\mathcal{F},i\in\mathcal{B},\label{eq:delayconst}\\
 & \quad\lim_{t\rightarrow\infty}\frac{\mathbb{E}\left[|Q_{f}^{i}|\right]}{t}=0,\forall f\in\mathcal{F},\forall i\in\mathcal{B},\label{eq:queueStability}\\
 & \mathbf{\quad\mathbf{x}}(t)\in\mathcal{R},\label{Rateconst}\\
 & \quad\boldsymbol{\pi}\in\Pi,\label{Piconst}\\
 & \quad\text{and }\:\eqref{eq:powerconstraint-0},\:\eqref{ratebounds},\nonumber
\end{align}
\end{subequations}where $\maximumdelay$ reflects the delay threshold required
for UEs, and $\delayprob \ll 1$ is the target probability for reliable
communication\footnote{The UEs can have different delay and reliability requirements.}. The probabilistic delay constraint \eqref{eq:delayconst}
implies that the probability that the delay for each flow at node
$i$ is greater than $\maximumdelay$ is very small, which captures the constraints
of ultra-low latency and reliable communication \textcolor{black}{\cite{vu2017ultra,elbamby2018toward}}.
It is also used to avoid congestion for each flow $f$ at any point
(BS) in the network, since the queue length is ensured less than $\maximumdelay\bar{\arrivalrate}_{f}$
with probability $1- \delayprob$. Hence, \eqref{eq:delayconst} forces
the transmission of all BSs without building large queues, and~\eqref{eq:queueStability}
maintains network stability.

The above problem has a non-linear probabilistic constraint \eqref{eq:delayconst},
which cannot be solved directly. Hence, we replace the non-linear
constraint \eqref{eq:delayconst} with a linear deterministic equivalent
by applying Markov's inequality \cite{2013queue,vu2017ultra} such
that $\text{Pr}\left(X\geq x\right)\leq\mathbb{E}\left[X\right]/x$
for a non-negative random variable $X$ and $x>0$. Thus, we relax
\eqref{eq:delayconst} as
\begin{equation}
\mathbb{E}\big[Q_{f}^{i}(t)\big]\leq \bar{\arrivalrate}_{f} \delayprob\maximumdelay.\label{appro_delayconst}
\end{equation}
Assuming that $\arrivalrate_{f}(t)$ follows a Poisson arrival process~\cite{2013queue},
we derive the expected queue length in \eqref{eq:queue1} for $i=0$ as
\begin{equation}
\mathbb{E}[Q_{f}^{i}(t)]  =t \bar{\arrivalrate}_{f}  -\sum_{\tau=1}^{t}\sum_{m=1,i_{f}^{({\rm o)}}\in\mathcal{Z}_{f}^{m}}\pi_{f}^{m}\ratemmWave^{(i,i_{f}^{({\rm o)}})}(\tau),\label{queue1-1}
\end{equation}
and the expected queue length in \eqref{eq:queue2}, for each SCBS, i.e.,
\begin{equation}
\mathbb{E}[Q_{f}^{i}(t)]=\sum_{\tau=1}^{t}\sum_{m}\pi_{f}^{m}\Big(\ratemmWave^{(i_{f}^{({\rm I)}},i)}\left(\tau\right)-\ratemmWave^{(i,i_{f}^{({\rm o)}})}(\tau)\Big).\label{queue2-1}
\end{equation}

\noindent Subsequently, combining the constraints \eqref{appro_delayconst}
and \eqref{queue1-1}, we obtain the following linear constraint \eqref{eq:delayconst01}
of instantaneous rate requirements, which helps to analyse and optimize the URLLC problem \cite{vu2017ultra,elbamby2018toward}, for MBS $i=0$,
\begin{equation}
\begin{alignedat}{1}
 \bar{\arrivalrate}_{f} (t-\delayprob\maximumdelay) & - \sum_{\tau=1}^{t-1}\sum_{m=1,i_{f}^{({\rm o)}}\in\mathcal{Z}_{f}^{m}}\pi_{f}^{m}\ratemmWave^{(i,i_{f}^{({\rm o)}})}(\tau)  \leq \\
 & \sum_{m=1,i_{f}^{({\rm o)}}\in\mathcal{Z}_{f}^{m}}\pi_{f}^{m}\ratemmWave^{(i,i_{f}^{({\rm o)}})}(t).
\end{alignedat}\label{eq:delayconst01}
\end{equation}

Similarly, for each SCBS $i=\{1,\cdots,B\}$, we have
\begin{equation}
\begin{alignedat}{1}
 -\bar{\arrivalrate}_{f} \delayprob\maximumdelay & +\sum_{\tau=1}^{t-1}\sum_{m}\pi_{f}^{m}\Big(\ratemmWave^{(i_{f}^{({\rm I)}},i)}\left(\tau\right)-\ratemmWave^{(i,i_{f}^{({\rm o)}})}\left(\tau\right)\Big) \leq  \\
& \sum_{m}\pi_{f}^{m}\Big(\ratemmWave^{(i,i_{f}^{({\rm o)}})}\left(t\right)-\ratemmWave^{(i_{f}^{({\rm I)}},i)}\left(t\right)\Big),
\end{alignedat}\label{eq:delayconst02}
\end{equation}
by combining \eqref{appro_delayconst} and \eqref{queue2-1}. With
the aid of the above derivations, we consider \eqref{eq:delayconst01}
and \eqref{eq:delayconst02} instead of \eqref{eq:delayconst} in
the original problem \eqref{OP1}. In practice, the statistical information
of all candidate paths to decide $\boldsymbol{\pi}_{f},\forall\,f\in\mathcal{F}$,
is not available beforehand, and thus solving \eqref{OP1} is challenging.
One solution is that paths are randomly assigned to each flow which
does not guarantee optimality, whereas applying an exhaustive search
is not practical. Therefore, in this work, the stochastic optimization is pertained to characterize the queuing latency in the presence of randomness (mmWave wireless channels and arbitrary arrivals). As a result, \eqref{OP1} is decoupled into sub-problems, which can be solved by low-complexity and efficient methods. In particular, RL is leveraged to find the best paths without requiring the statistic information, and SCA method obtains a locally efficient solution for assigning rate over the flows.
\section{Proposed Path Selection and Rate Allocation Algorithm }
\label{Design}
In this section, we propose a Lyapunov stochastic optimization based framework to solve our predefined problem \eqref{OP1} with relaxed latency constraints. To do that, we first introduce a set of auxiliary variables to refine the original problem \eqref{OP1}. Next, we convert the constraints into virtual queues and write the conditional Lyapunov drift function. Finally, the solution of the equivalent problem is obtained by minimizing the Lyapunov drift and a penalty from the objective function. Let us start by rewriting \eqref{OP1} equivalently as follows:
\begin{subequations}\label{LP2}
\begin{eqnarray}
\ \mbox{RP:}\,\max_{\bar{\boldsymbol{\varphi}},\boldsymbol{\pi},{\bf p}} &  & \overline{U_{0}(\boldsymbol{\varphi})}\label{LP1:obj}\\
\text{subject to} &  & \bar{\varphi}_{f}-\bar{x}_{f}\leq0,\ \forall f\in{\cal F},\label{lp1:setconstraint}\\
 &  & \eqref{eq:powerconstraint-0},\:\eqref{ratebounds},\,\eqref{eq:queueStability},\,\eqref{Piconst},\,\eqref{eq:delayconst01},\,\eqref{eq:delayconst02},\nonumber
\end{eqnarray}
\end{subequations}where the new constraint \eqref{lp1:setconstraint}
is introduced to replace the rate constraint \eqref{Rateconst} with
new auxiliary variables $\boldsymbol{\varphi}=\left(\varphi_{1},\:\cdots,\varphi_{F}\right)$.
In \eqref{lp1:setconstraint}, $\bar{\boldsymbol{\varphi}}\triangleq\lim\limits _{t\to\infty}\frac{1}{t}\sum_{\tau=0}^{t-1}\mathbb{E}\left[|\boldsymbol{\varphi}(\tau)|\right]$.
In order to ensure the inequality constraint \eqref{lp1:setconstraint},
we introduce a virtual queue vector $Y_{f}\left(t\right),$ which
is given by

\begin{equation}
Y_{f}\left(t+1\right)=\left[Y_{f}\left(t\right)+\varphi_{f}\left(t\right)-x_{f}\left(t\right)\right]^{+},\:\forall f\in{\cal F}.\label{eq:virtualQ}
\end{equation}
\textcolor{black}{Let $\queuebacklog(t)=({\bf Q}(t),\,{\bf Y}(t))$ denote
the queue backlogs. We first write the conditional Lyapunov drift for slot $t$ as}

\begin{equation}
\lyapunovdrift =\mathbb{E}\Big[L\left(\queuebacklog(t+1)\right)-L\left(\queuebacklog(t)\right)|\queuebacklog(t)\Big],\label{eq:drift}
\end{equation}
where $L\Big(\queuebacklog(t)\Big)\triangleq\frac{1}{2}\Big[\sum_{f=1}^{F}\sum_{i=0}^{B}Q_{f}^{i}(t)^{2}+\sum_{f=1}^{F}Y_{f}(t)^{2}\Big]$
is the quadratic Lyapunov function of $\queuebacklog(t)$
\cite{neely2010S}. By applying the Lyapunov drift-plus-penalty
technique \cite{Vu_LB, neely2010S}, the solution
of \eqref{LP2} is obtained by minimizing the Lyapunov drift and a penalty from the objective function, i.e.,

\begin{eqnarray}
\min &  & \lyapunovdrift - \controlpara \mathbb{E}\left[U_{0}\left(\boldsymbol{\varphi}\right)|\queuebacklog(t)\right].\label{eq:driftplusPenalty}
\end{eqnarray}
Here, $\controlpara$ is a control parameter to trade off utility optimality
and queue length \cite{Vu_LB, neely2010S}. Moreover, the stability of $\queuebacklog(t)$
ensures that the constraints of problem \eqref{eq:queueStability}
and \eqref{lp1:setconstraint} are held. Noting that $\big([a]^{+}\big)^2 \leq  a^2$ and $(a \pm b)^2 \leq a^2 \pm 2ab + b^2$
for any real positive number $a, b$, and thus, by neglecting other indexes $t, f,  \ldots$, we have the following inequalities
\begin{align}
\big([{Q} - R^{({\rm o)}}]^{+} + R^{({\rm I)}} \big)^{2} - {Q}^2 & \leq 2 {Q}(R^{({\rm I})}- R^{(\rm o)}) + ( R^{ (\rm I)}- {R}^{(\rm o)} )^{2}, \notag \\
\big([{Q} - R^{({\rm o)}}] + a \big)^{2} - {Q}^2 & \leq 2 {Q}(a- R^{(\rm o)}) + ( a- {R}^{(\rm o)} )^{2}, \notag \\
\big([{Y} + \varphi- {x}]^{+}\big)^2 - {Y}^2 & \leq 2 {Y}(\varphi- {x}) + (\varphi- {x})^{2}. \notag
\end{align}
\noindent Subsequently, following the calculations of the Lyapunov optimization \cite{neely2010S}, choosing that $\boldsymbol{\varphi}\in{\cal R}$
and a feasible ${\bf \pi}$ and all possible $\queuebacklog(t)$ for all $t$, we obtain
\begin{eqnarray}
\eqref{eq:driftplusPenalty}
& \leq & \sum_{f=1}^{F}\sum_{i=1}^{B}Q_{f}^{i}\,\mathbb{E} \Big[\sum_{m}\pi_{f}^{m} (\ratemmWave^{(i_{f}^{({\rm I)}},i)}-\ratemmWave^{(i,i_{f}^{({\rm o)}})} )|\queuebacklog(t) \Big] \nonumber \\
 &  & -\sum_{f=1}^{F}Q_{f}^{i|i=0}\:\mathbb{E}\Big[\sum_{m=1,i_{f}^{({\rm o)}}\in\mathcal{Z}_{f}^{m}}\pi_{f}^{m}\ratemmWave^{(i,i_{f}^{({\rm o)}})}|\queuebacklog(t)\Big]\label{eq:drift2}\\
 &  & +\sum_{f=1}^{F}\mathbb{E}\Big[Y_{f}\varphi_{f}-\nu U\left(\varphi_{f}\right)-Y_{f}x_{f}|\queuebacklog(t)\Big]+\Psi.\nonumber
\end{eqnarray}
\textcolor{black}{Here, $\Psi$  is a finite constant
that satisfies $\Psi \geq \frac{1}{2}\sum_{f=1}^{F}\sum_{i=1}^{B}\mathbb{E} \big[\sum_{m}\pi_{f}^{m} (\ratemmWave^{(i_{f}^{({\rm I)}},i)}-\ratemmWave^{(i,i_{f}^{({\rm o)}})} )^2|\queuebacklog(t) \big] + \frac{1}{2} \sum_{f=1}^{F} \mathbb{E} \big[\sum_{m=1,i_{f}^{({\rm o)}}\in\mathcal{Z}_{f}^{m}}\pi_{f}^{m} ( \arrivalrate_f  - \ratemmWave^{(i,i_{f}^{({\rm o)}})} )^2|\queuebacklog(t) \big] + \frac{1}{2} \sum_{f=1}^{F}\mathbb{E} \big[ (\varphi_{f}-x_{f})^2|\queuebacklog(t)\big]$ \cite{neely2010S,Vu_LB}. }The solution to
\eqref{LP2} can be obtained by minimizing the upper bound in \eqref{eq:drift2}.
For every slot $t,$ observing $\queuebacklog(t)$, we
have three decoupled subproblems and provide the solutions for each
subproblem as follows. The flow-split vector and the probability distribution are determined by

\begin{align*}
\mbox{SP1}:\,\min_{\boldsymbol{\pi}} & \qquad\sum_{f=1}^{F} \flowutility_{f}\\
\text{subject to} & \qquad\eqref{Piconst},
\end{align*}
where

\begin{equation}
\begin{alignedat}{1}
\flowutility_{f}  =  & \sum_{i=1}^{B}Q_{f}^{i}\sum_{m}\pi_{f}^{m}\left(\ratemmWave^{(i_{f}^{({\rm I)}},i)}-\ratemmWave^{(i,i_{f}^{({\rm o)}})}\right) \\ \notag
 &-Q_{f}^{i|i=0}\sum_{m=1,i_{f}^{({\rm o)}}\in\mathcal{Z}_{f}^{m}}\pi_{f}^{m}R_{(i,i_{f}^{({\rm o)}})}^{f}. \notag
\end{alignedat}{1}
\end{equation}
Then, we select the optimal auxiliary variables by solving

\begin{eqnarray*}
\mbox{SP2:}\:\min_{\boldsymbol{\varphi}|\boldsymbol{\pi}} &  & \sum_{f=1}^{F}\Big[Y_{f}\,\varphi_{f}-\nu U\left(\varphi_{f}\right)\Big]\\
\mbox{\mbox{subject to}} &  & \varphi_{f}(t)\geq0,\:\forall f\in{\cal F}.
\end{eqnarray*}
Let $\varphi_{f}^{\ast}$ be the optimal solution obtained by the
first order derivative of the objective function of SP2. Assuming
a logarithmic utility function, we have $\varphi_{f}^{\ast}(t)=\max\left\{ \frac{\nu}{Y_{f}},\:0\right\} .$
Finally, the \textbf{RA} is done by assigning transmit power,
which is obtained by

\begin{eqnarray*}
\mbox{SP3:}\:\min_{{\bf x},{\bf p}|\boldsymbol{\pi}} &  & \sum_{f=1}^{F}-Y_{f}\,x_{f}\\
\text{subject to} &  & \eqref{eq:powerconstraint-0},\:\eqref{ratebounds},\,\eqref{eq:delayconst01},\,\eqref{eq:delayconst02}.
\end{eqnarray*}
\vspace{-3em}
\subsection{Path Selection}
Recall that ${\bf z}_{f}\,$ represents the flow-split vector given to flow $f\,$ and $z_{f}^{m}=1$ when path $m$ is used to send data for flow $f$. The MBS selects paths for each flow with a given probability (mixed strategy) \cite{2011learning}. We denote $u_{f}^{m}=u_{f}\left(z_{f}^{m},{\bf z}_{f}^{-m}\right)$ as a utility function of flow $f$ when using path $m$. The vector ${\bf z}_{f}^{-m}$ denotes the flow-split vector excluding path $m$. The MBS can choose more than one path to deliver data, from $\mbox{SP1}$, the utility gain of flow $f$ is
\vspace{-0.5em}
\[
u_{f}=\sum_{m}u_{f}^{m}=-\flowutility_{f}.
\]
\vspace{-0.5em}

\noindent To exploit the historical information, the MBS determines a flow-split vector for each flow $f$ from ${\cal Z}_{f}$ based on the PMF from the previous stage $t-1$, i.e.,\vspace{-1em}
\begin{equation}
{\bf \boldsymbol{\pi}}_{f}\left(t-1\right)=\left(\pi_{f}^{1}\left(t-1\right),\cdots,\pi_{f}^{Z_{f}}\left(t-1\right)\right).\label{eq:PreviousProbability}
\end{equation}
Here, we define ${ \boldsymbol \Phi}_{f}(t)=(\Phi_{f}^{1}\left(t\right),\cdots,\Phi_{f}^{m}\left(t\right)\cdots,\Phi_{f}^{Z_{f}}\left(t\right))$
as a regret vector of determining flow-split vector for flow $f$. The MBS selects the flow-split vector with highest regret in which
the mixed-strategy probability is given as
\vspace{-0.5em}
\begin{equation}
\pi_{f}^{m}\left(t\right)=\frac{\left[\Phi_{f}^{m}\left(t\right)\right]^{+}}{\sum_{m'\in\mathcal{Z}_{f}}\left[\Phi_{f}^{m'}\left(t\right)\right]^{+}}.\label{eq:highestregret}
\end{equation}
Let $\hat{{\boldsymbol \Phi}}_{f}(t)=(\hat{\Phi}_{f}^{1}\left(t\right),\cdots,\hat{\Phi}_{f}^{m}\left(t\right)\cdots,\hat{\Phi}_{f}^{Z_{f}}\left(t\right))$
be the estimated regret vector of flow $f$. \textcolor{black}{Basically, with the goal of maximizing the cumulative reward in SP1, the MBS (agent) has to discover the possible paths (action set) in order to find the best paths (distribution of actions with higher pay-off) in the long run \cite{2011learning}. If the MBS spends much time on discovering paths (called exploration), it leads to longer convergence time. If the MBS only exploits an action (called exploitation), which gave the highest pay-off at the beginning, it may loose a chance to obtain higher reward later. Hence, balancing the trade-off between exploration and exploitation is fundamental of efficient learning. For these purpose, we have adopted the logit of Boltzmann-Gibbs (BG) kernel to efficiently learn the best paths \cite{2011learning, bennis2013self}}, $\boldsymbol{\boltzmanfactor}_{f}^{m}\left(\hat{{\boldsymbol \Phi}}_{f}(t)\right)$, given by
\begin{equation}
\begin{alignedat}{1}
\boldsymbol{\boltzmanfactor}_{f}^{m}\left(\hat{{\bf \Phi}}_{f}(t)\right)\:=\: & \underset{\boldsymbol{\pi}_{f}\in\Pi}{\mbox{argmax}}\sum_{m\in{\cal Z}_{f}}\left[\pi_{f}^{m}\left(t\right)\hat{\Phi}_{f}^{m}\left(t\right)\right.\\
 & \qquad\quad\left.-\kappa_{f}\pi_{f}^{m}\left(t\right)\ln(\pi_{f}^{m}\left(t\right))\right],
\end{alignedat}
\label{eq:GibbsDistribution}
\end{equation}
where the trade-off factor $\kappa_{f}$ is used to balance between
exploration and exploitation \cite{2012Perlaza,bennis2013self,2013backhaul}.\textcolor{black}{{}
If $\kappa_{f}$ is small, the MBS selects ${\bf z}_{f}$ with highest
payoff. For $\kappa_{f}\rightarrow\infty$ all decisions have equal
probability.}

For a given set of $\hat{{\boldsymbol \Phi}}_{f}(t)$ and $\kappa_{f}$, we solve
\eqref{eq:GibbsDistribution} to find the probability distribution
in which the solution determining the disjoint paths for each flow
$f$ is given as
\begin{equation}
\boltzmanfactor_{f}^{m}(\hat{{\boldsymbol \Phi}}_{f}(t))=\frac{\exp\left(\frac{1}{\kappa_{f}}\left[\hat{\Phi}_{f}^{m}\left(t\right)\right]^{+}\right)}{\sum\limits _{m'\in{\cal Z}_{f}}\exp\left(\frac{1}{\kappa_{f}}\left[\hat{\Phi}_{f}^{m'}\left(t\right)\right]^{+}\right)}.\label{eq:GibbsSolution}
\end{equation}
We denote $\hat{u}\left(t\right)$ as the estimated utility of flow
$f$ at time instant $t$ with action ${\bf z}_{f}$, i.e, $\hat{{\bf u}}_{f}(t)=(\hat{u}_{f}^{1}\left(t\right),\cdots,\hat{u}_{f}^{m}\left(t\right)\cdots,\hat{u}_{f}^{Z_{f}}\left(t\right))$.
Upon receiving the feedback, $\tilde{u}_{f}(t)$ denotes the utility
observed by flow $f$, i.e., $\tilde{u}_{f}(t)=u_{f}(t-1)$, we propose
the learning mechanism at each time instant $t$ as follows.

\textbf{\textit{\textcolor{black}{Learning procedure}}}: The estimates
of the utility, regret, and probability distribution functions are
performed, and are updated for all actions per path $m$ as follows:
\vspace{-0.5em}
\begin{equation}
\begin{cases}
\hat{u}_{f}^{m}\left(t\right)=\hat{u}_{f}^{m}\left(t-1\right)+\learningratei(t)\mathbb{I}_{\{{\bf z}_{f}={\bf z}_{f}^{m}\}}\Bigl(\tilde{u}_{f}(t)-\hat{u}_{f}^{m}\left(t-1\right)\Bigr),\\
\hat{\Phi}_{f}^{m}\left(t\right)=\hat{\Phi}_{f}^{m}\left(t-1\right)+\learningrateii(t)\left(\hat{u}_{f}^{m}(t)-\tilde{u}_{f}(t)-\hat{\Phi}_{f}^{m}\left(t-1\right)\right),\\
\pi_{f}^{m}(t)=\pi_{f}^{m}(t-1)+\learningrateiii(t)\left(\boltzmanfactor_{f}^{m}(\hat{{\boldsymbol \Phi}}_{f}(t))-\pi_{f}^{m}(t-1)\right),
\end{cases}\label{eq:Probability}
\end{equation}
 Here, $\learningratei(t)$, $\learningrateii(t)$, and $\learningrateiii(t)$ are the
learning rates (please see Section \ref{performance_analysis}
for more details and convergence proof). Based on the probability
distribution as per \eqref{eq:Probability}, the MBS determines the
flow-split vector for each flow $f$. The learning-aided
\textbf{PS} is performed in a long-term period to ensure that the
paths do not suddenly change, and thus, the SCBSs have sufficient time
to deliver data. For instance, at the beginning
of the large time scale, the best paths are selected, and will be
used for the rest of these large scale time slots as shown in Fig.
\ref{Flowchart}.

\vspace{-1em}

\subsection{\textcolor{black}{Rate Allocation}}

Consider $\ratemmWave^{(i,j)}=\log(1+p_{(i,j)}^{f}|g_{(i,j)}({\bf h})|^{2})$
as the transmission rate, where the effective channel gain\footnote{The effective channel gain captures the path loss, channel variations,
and interference penalty (Here, the impact of interference is considered
small due to highly directional beamforming and high pathloss for
interfered signals at mmWave frequency band, and thus a multi-hop directional transmission can be operated at dense mmWave networks
\cite{2015tractable,2015transit,2017INR,2015analysis,2016regimes}). } for mmWave channels can be modeled as $|g_{(i,j)}({\bf h})|^{2}=\frac{|\tilde{g}_{(i,j)}({\bf h})|^{2}}{1+I^{\max}}$
\cite{2013beamforming,Vu_LB}. Here, $\tilde{g}_{(i,j)}({\bf h})$
and $I^{\max}$ denote the normalized channel gain and the maximum
interference, respectively. Denoting the left hand side (LHS) of \eqref{eq:delayconst01}
and \eqref{eq:delayconst02} as $D_{i}^{f}$ for simplicity, the optimal
values of flow control ${\bf x}$ and transmit power ${\bf p}$ \textcolor{black}{in
the sub-problem 3 (SP3)} are found by minimizing

\vspace{-1em}\begin{subequations}
{\small{}\label{RateAllocation} }
\begin{alignat}{1}
\min_{{\bf x},{\bf p}|\boldsymbol{\pi}} & \quad\sum_{f=1}^{F}-Y_{f}x_{f}\label{eq:rateallocation}\\
\text{subject to} & \quad1+p_{(i,i_{f}^{({\rm o)}})}^{f}|g_{(i,i_{f}^{({\rm o)}})}|^{2}\geq e^{x_{f}},\forall f\in{\cal F},\,i=0,\label{eq:excontraint1}\\
 & \quad\frac{1+p_{(i,i_{f}^{({\rm o)}})}^{f}|g_{(i,i_{f}^{({\rm o)}})}|^{2}}{1+p_{(i_{f}^{({\rm I)}},i)}^{f}|g_{(i_{f}^{({\rm I)}},i)}|^{2}}\geq e^{D_{i}^{f}},f\in{\cal F},\forall i=1:B,\label{eq:excontraint2}\\
 & \quad\sum_{f\in F}p_{(i,i_{f}^{({\rm o)}})}^{f}\leq P_{i}^{\text{max}},\forall i\in{\cal B},\forall f\in{\cal F}.\label{eq:powerconstraint}
\end{alignat}
\end{subequations}{
\noindent The constraint \eqref{eq:excontraint2} is non-convex, motivated by the low-complexity of SCA method, we solve \eqref{RateAllocation} by replacing \eqref{eq:excontraint2} with its proper convex approximation \cite{2015nguyen, Vu_LB, Vu2016}. Since it is very hard to find the convex approximation of \eqref{eq:excontraint2}   \cite{beck2010seq, 2004convex}, we introduce the slack variable $y$ to transform \eqref{eq:excontraint2} into equivalent constraints, which having a proper bound satisfying the conditions in \cite[Property A]{beck2010seq} as
\hspace{-1.5em}
\begin{alignat}{1}
\frac{2+p_{(i,i_{f}^{({\rm o)}})}^{f}|g_{(i,i_{f}^{({\rm o)}})}|^{2}}{2} & \geq\sqrt{y^{2}+\Big(\frac{p_{(i,i_{f}^{({\rm o)}})}^{f}|g_{(i,i_{f}^{({\rm o)}})}|^{2}}{2}\Big)^{2}},\label{eq:cone1}\\
\frac{y^{2}}{1+p_{(i_{f}^{({\rm I)}},i)}^{f}|g_{(i_{f}^{({\rm I)}},i)}|^{2}} & \geq e^{D_{i}^{f}}.\label{eq:cone2}
\end{alignat}
Here, the constraint \eqref{eq:cone1} holds a form of the second-order cone inequalities \cite{2004convex, beck2010seq, 2001lectures}, while the LHS of
constraint \eqref{eq:cone2} is a quadratic-over-affine function which is iteratively replaced by the first order to achieve a convex approximation as follows:
\begin{alignat}{1}
\frac{2yy^{(l)}}{1+p_{(i_{f}^{({\rm I)}},i)}^{f(l)}|g_{(i_{f}^{({\rm I)}},i)}|^{2}}-\frac{y^{(l)2}\Big(1+p_{(i_{f}^{({\rm I)}},i)}^{f}|g_{(i_{f}^{({\rm I)}},i)}|^{2}\Big)}{\Big(1+p_{(i_{f}^{({\rm I)}},i)}^{f(l)}|g_{(i_{f}^{({\rm I)}},i)}|^{2}\Big)^{2}}\geq e^{D_{i}^{f}}.\label{eq:cone3}
\end{alignat}
Here, the superscript $l$ denotes the $l$th iteration. Hence, we
iteratively solve the approximated convex problem of \eqref{RateAllocation}
as \textbf{Algorithm}~\ref{algRate1} in which the approximated problem\footnote{Note that the problem of finding the global optimality is outside
the scope of our study. The effectiveness of SOCP method was verified
in the literature and shown to be robust in practical scenarios \cite{2004convex}.} is given as
\begin{eqnarray}
\min_{{\bf x},{\bf p}|\boldsymbol{\pi}} &  & \sum_{f=1}^{F}-Y_{f}x_{f}\label{Optimal-Rate}\\
\text{subject to} &  & \eqref{ratebounds}, \, \eqref{eq:powerconstraint},\,\eqref{eq:excontraint1},\,\eqref{eq:cone1},\,\eqref{eq:cone3}.\nonumber
\end{eqnarray}
\begin{algorithm}
\label{algRate}\begin{algorithmic}

\STATE Initialization: set $l=0$ and generate initial points $y^{(l)}$.

\REPEAT \STATE $\text{Solve}$~(\ref{Optimal-Rate}) with $y^{(l)}$
to get the optimal value $y^{(l)\star}$.

\STATE $\text{Update}$ $y^{(l+1)}:=y^{(l)\star}$; $l:=l+1$.

\UNTIL{\text{Convergence}}

\end{algorithmic} \caption{Iterative \textbf{RA}}
\label{algRate1}
\end{algorithm}
Finally, the information flow diagram of the learning-aided \textbf{PS}
and \textbf{RA} approach is shown in Fig. \ref{Flowchart}, where
the \textbf{RA} is executed in a short-term period. Note that
the \textbf{PS} and \textbf{RA} are both done at the MBS, in
this work we assume that the information is shared among the base
stations by using the X2 interface. As opposed to
a brute-force approach yielding the global optimal solution, the proposed
iterative solution that uses time scale separation remarkably reduces
the search time and computational complexity, while obtaining an efficient
suboptimal solution.
\begin{figure}
\centering\includegraphics[width=1\columnwidth]{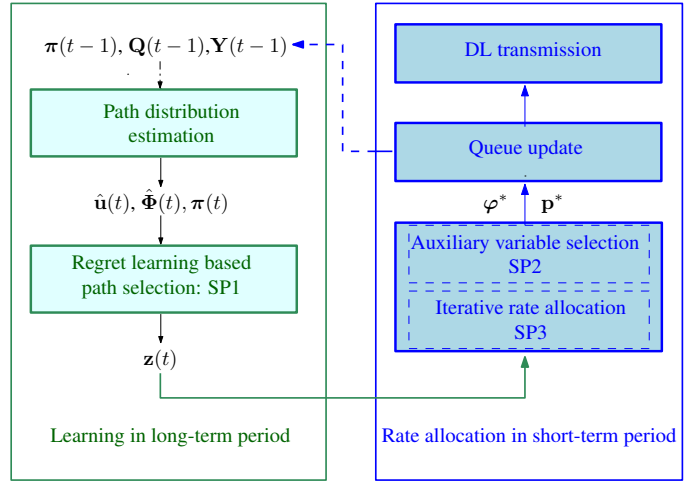}
\caption{Information flow diagram of the learning-aided \textbf{PS} and
\textbf{RA} approach.}
\label{Flowchart}
\end{figure}
\vspace{-1.5em}
\section{Performance Analysis}
\label{performance_analysis}

In this section, we provide a comprehensive performance analysis of
our proposed Lyapunov optimization based framework. We show that there
exists an $[\mathcal{O}(1/\nu),\,\mathcal{O}(\nu)]$ utility-queue
backlog trade-off, where $\nu$ is the Lyapunov control parameter
\cite{neely2010S}. Next, we present the conditions that ensure that
the proposed learning-based \textbf{PS} converges with probability
one. Finally, a convergence analysis and a complexity computation
of the SOCP based approximation method for \textbf{RA} sub-problem
are studied.
\vspace{-1em}
\subsection{Queue and Utility Performance}
\label{Analysis1}
We scrutinize the performance analysis of our proposed algorithm and
prove that the queues are stable as per the following theorem.
\begin{theorem}
\label{Theo1}{[}Optimality{]} Assume that all queues are initially
empty. For arbitrary arrival rates, the \textbf{PS} and \textbf{RA}
are chosen to satisfy~(\ref{eq:drift2}) and the rate regime. For
a given constant $\chi\geq0$, the network utility maximization with
any $\nu>0$ provides the following utility performance with $\chi-\textit{approximation}$
\[
U_{0}\geq U_{0}^{\ast}-\frac{\Psi+\chi}{\nu},
\]
where $U_{0}^{\star}$ is the optimal network utility over the rate
regime.
\end{theorem}
\begin{IEEEproof} We first prove the queues are bounded. Let $\varkappa_{f}$
denote the largest right derivative of $U\left(\bar{x}_{f}\right)$,
the Lyapunov framework can guarantee the following strong stability
of the virtual queues and the network queues as follows
\begin{equation}
Q_{f}^{i}(t)\leq\nu\varkappa_{f}+\maximumrate_f,\label{Bound1}
\end{equation}
\begin{equation}
Y_{f}(t)\leq\nu\varkappa_{f}+\maximumrate_f.\label{Bound2}
\end{equation}
Here, we first prove the bound of the virtual queues, and then the
bound of the network queues are proved similarly. Suppose that all
queues are initially empty at $t=1$, this clearly holds for $t=1$.
Suppose these inequalities hold for some $t>1$, we need to show that
it also holds for $t+1$.

From~(\ref{eq:virtualQ}), if $Y_{f}(t)\leq\nu\varkappa_{f}$ then
$Y_{f}(t+1)\leq\nu\varkappa_{f}+\maximumrate_f$ and the bound holds
for $t+1$ due to the rate constraint $x_{f}(t)\leq \maximumrate_f$.
Else, if $Y_{f}(t)\geq\nu\varkappa_{f}$; since the value of auxiliary
variables is determined by maximized ${\textstyle \sum_{f=1}^{\emph{F}}Y_{f}\left(t\right)\varphi_{f}\left(t\right){\textstyle -\nu U_{0}\left(\boldsymbol{\varphi}\left(t\right)\right)}}$,
$\boldsymbol{\varphi}\left(t\right)$ is then forced to be zero. From~(\ref{eq:virtualQ}),
$Y_{f}(t+1)$ is bounded by $Y_{f}(t)$. Since the virtual queues
are bounded for $t$, we have the following inequalities
\begin{equation}
Y_{f}(t+1)\leq Y_{f}(t)\leq \nu \varkappa_{f} + \maximumrate_f,\label{Bound22}
\end{equation}
Hence, the bounds of the virtual queues hold for all $t$. Similarly,
we can prove that the network queue~(\ref{Bound1}) is stable with
network queues in (\ref{eq:queue1}) and (\ref{eq:queue2}).

We have established the network bounds, we are now going to show the
utility bound. Since our solution of (\ref{OP1}) is to minimize the
Lyapunov drift and the objective function every time slot $t$, we
have the following inequality given all existing $\queuebacklog(t)$
for all $t$,
\[
\begin{aligned}  &\mathbf{\Delta}(\mathbf{\queuebacklog}(t))-\nu\mathbb{E}\left[U_{0}(\boldsymbol{\bar{\varphi}}(t))|\queuebacklog(t)\right]\leq \\
 & \sum_{f=1}^{F}\sum_{i=1}^{B}Q_{f}^{i}\,\mathbb{E}\Big[\sum_{m}\pi_{f}^{m\ast}\left(\ratemmWave^{(i_{f}^{({\rm I)}},i)}-\ratemmWave^{(i,i_{f}^{({\rm o)}})}\right)|\queuebacklog(t)\Big] \\
 & -\sum_{f=1}^{F}Q_{f}^{i|i=0}\:\mathbb{E} \Big[\sum_{m=1,i_{f}^{({\rm o)}}\in\mathcal{Z}_{f}^{m}}\pi_{f}^{m\ast}\ratemmWave^{(i,i_{f}^{({\rm o)}})}\Big|\queuebacklog(t)\Big]  \\
 &+\sum_{f=1}^{F}\mathbb{E}\Big[Y_{f}\varphi_{f}^{\ast} -\nu U\left(\varphi_{f}^{\ast}\right)-Y_{f}x_{f}^{\ast}|\queuebacklog(t)\Big]+\Psi,
\end{aligned}
\]
where $\boldsymbol{\varphi}_{f}^{\ast}\left(t\right)$, and $x_{f}^{\ast}\left(t\right)$
are the optimal values of the sub-problems SP2 and SP3, respectively.
Here, $\pi_{f}^{m\ast}$ and $z_{f}^{m\ast}$ are the optimal values
of the sub-problem SP1. Since the queues are bounded, for a given
$\chi\geq0$, we obtain
\vspace{-0.5em}
\begin{equation}
\begin{split}\mathbf{\Delta}(\mathbf{\queuebacklog}(t)) & -\nu\mathbb{E}\left[U_{0}(\boldsymbol{\varphi}(t))|\queuebacklog(t)\right]
\leq \\ \Psi &-\nu\mathbb{E}\left[U_{0}(\boldsymbol{\varphi}^{\ast}(t))|\queuebacklog(t)\right]+\chi.
\end{split}
\label{Proof2}
\end{equation}
By taking expectations of both sides of the above inequality and choosing
$\mathbf{x}^{\ast}\left(t\right)=\boldsymbol{\varphi}^{\ast}\left(t\right)$,
it yields for all $t\geq0$,
\begin{equation}
\begin{split}
\mathbb{E}\left[L(\queuebacklog(t+1))-L(\mathbf{\queuebacklog}(t))|\mathbf{\queuebacklog}(t)\right]  &-\nu\mathbb{E}\left[U_{0}(\boldsymbol{\varphi}(t))|\queuebacklog(t)\right]\leq \\
& \Psi+\chi-\nu\mathbb{E}\left[U_{0}(\mathbf{x}^{\ast}(t)).\right]
\end{split}
\label{Proof3}
\end{equation}
By taking the sum over $\tau=1,\ldots,t$ and dividing by $t$, (using
the fact that $U_{0}\left(\mathbf{x}^{\ast}\left(t\right)\right)=U_{0}^{\ast}$),
yielding
\begin{equation}
\begin{split}
& \frac{\mathbb{E}\big[L(\mathbf{\queuebacklog}(t+1))-L(\mathbf{\Sigma}(0))|\mathbf{\queuebacklog}(t)\big]}{t} \\
& -\frac{\nu}{t}\sum_{\tau=1}^{t}\mathbb{E}\left[U_{0}(\boldsymbol{\varphi}(t))|\queuebacklog(t)\right] \leq  ~\Psi+\chi-\nu U_{0}^{\ast}.
\label{Proof4}
\end{split}
\end{equation}
By using the fact that $L\left(\mathbf{\queuebacklog}\left(t+1\right)\right)\geq0$
and $L\left(\queuebacklog\left(\tau=1\right)\right)=0$,
and applying Jensen's inequality in the concave function and rearranging
the terms yields
\[
U_{0}(\boldsymbol{\varphi}(t))\geq~U_{0}^{\ast}-\frac{\Psi+\chi}{\nu}.
\]
Since the network utility function is a non-decreasing concave function,
the auxiliary variable is chosen to satisfy $x_{f}(t)\geq\varphi_{f}(t)$.
Hence $U_{0}\left(\mathbf{x}\left(t\right)\right)\geq U_{0}(\boldsymbol{\varphi}(t))\geq~U{}_{0}^{\ast}-\frac{\Psi+\chi}{\nu}$,
which means that the solution is closed to the optimal as increasing
$\nu$. Which completes the proof of the $\textbf{Theorem~\ref{Theo1}}$.
\end{IEEEproof}

Hence, there exists an $[\mathcal{O}(\nu),\:\mathcal{O}(1/\nu)]$
utility-queue length trade-off, which leads to an utility-delay balancing.
We now prove that all queues are stable, the bound (\ref{Proof4})
can be rewritten as
\[
\mathbf{\Delta}(\mathbf{\queuebacklog}(t))\leq~C,
\]
where $C$ is any constant that satisfies for all $t$ and $\mathbf{\queuebacklog}(t)$:
$C\geq\Psi+\chi-\nu(U_{0}^{\ast}-\mathbb{E}[U_{0}(\boldsymbol{\varphi}(t))|\queuebacklog(t)])$.
By using the definition of the Lyapunov drift and taking an expectation,
obtaining
\[
\mathbb{E}\big[L(\mathbf{\queuebacklog}(t))\big]\leq~Ct.
\]
As the definition of the Lyapunov function $L(\mathbf{\queuebacklog}(t))$,
$\forall i\in{\cal B}$ we have
\[
\mathbb{E}[Q_{f}^{i}(t)]^{2},\mathbb{E}[Y_{f}(t)]^{2}\leq~2Ct.
\]
Dividing both sides by $t^{2}$, and taking the square roots shows
for all $t>0$ and $\forall i\in{\cal B}$:
\[
\frac{\mathbb{E}[Q_{f}^{i}(t)]}{t},\frac{\mathbb{E}[Y_{f}(t)]}{t}\leq~\sqrt{\frac{2C}{t}}.
\]
As $t\rightarrow\infty$, taking the limit, we prove the queues are
stable.

\vspace{-1em}

\subsection{Learning Convergence Conditions}
\label{Analysis2}
Due to the space limitation, the complete convergence conditions can be found in \cite{bennis2013self}. Here, we briefly establish the convergence conditions to the $\text{\textopeno}$-coarse
correlated equilibrium for the reinforcement learning based algorithm,
where $\text{\textopeno}$ is a very small positive value  \cite{2000convergence}.
The complete proof was studied in \cite{bennis2013self,2013backhaul},
the learning rates $\learningratei\left(t\right)$, $\learningrateii\left(t\right)$,
and $\learningrateiii\left(t\right)$ are chosen to satisfy the convergence
conditions as follows:
\begin{alignat*}{1}
{\textstyle {\textstyle }\begin{cases}
\underset{t\rightarrow\infty}{\lim}\sum_{\tau=0}^{t}\learningratei\left(\tau\right)=+\infty, & \underset{t\rightarrow\infty}{\lim}\sum_{\tau=0}^{t}\learningrateii\left(\tau\right)=+\infty,\\
\underset{t\rightarrow\infty}{\lim}\sum_{\tau=0}^{t}\learningrateiii\left(\tau\right)=+\infty, & \underset{t\rightarrow\infty}{\lim}\sum_{\tau=0}^{t}\learningrateis\left(\tau\right)<+\infty,\\
\underset{t\rightarrow\infty}{\lim}\sum_{\tau=0}^{t}\learningrateiis\left(\tau\right)<+\infty, & \underset{t\rightarrow\infty}{\lim}\sum_{\tau=0}^{t}\learningrateiiis\left(\tau\right)<+\infty,\\
\underset{t\rightarrow\infty}{\lim}\frac{\learningrateiii\left(t\right)}{\learningrateii\left(t\right)}=0, & \underset{t\rightarrow\infty}{\lim}\frac{\learningrateii\left(t\right)}{\learningratei\left(t\right)}=0.
\end{cases}}
\end{alignat*}
\vspace{-1em}

\subsection{Convergence Analysis of SOCP based $\textbf{Algorithm~\ref{algRate1}}$}

\label{Analysis3}

We establish a convergence result for $\textbf{Algorithm~\ref{algRate1}}$
based on the SOCP approach. By using the SOCP approach, we have approximated
the original non-convex problem~(\ref{RateAllocation}) by a strongly
convex problem~(\ref{Optimal-Rate}). We briefly describe the convergence
for the sake of completeness since it was studied in~\cite{beck2010seq}.
We assume that the $\textbf{Algorithm~\ref{algRate1}}$ obtains the
solution of problem~(\ref{Optimal-Rate}) at iteration $l+1~\mathrm{th}$.
The updating rule in $\textbf{Algorithm~\ref{algRate1}}$ ensures
that the optimal values $\mathbf{y}^{(l)}$ at iteration $l$ satisfy
all constraints in~(\ref{Optimal-Rate}) and are feasible to the
optimization problem at iteration $l+1$. Therefore, the objective
obtained in the $l+1\mathrm{st}$ iteration is less than or equal
to that in the in the $l\mathrm{th}$ iteration, since we minimize
the linear function. In other words, $\textbf{Algorithm~\ref{algRate1}}$
yields a non-increasing sequence. Due to the transmit power constraints
and rate constraints, the objective is bounded, and thus $\textbf{Algorithm~\ref{algRate1}}$
converges to some local optimal solution of~(\ref{Optimal-Rate}).
Moreover, $\textbf{Algorithm~\ref{algRate1}}$ produces a sequence
of points that are feasible for the original problem~(\ref{RateAllocation})
and this solution satisfies the Karush\textendash Kuhn\textendash Tucker
(KKT) condition of the original problem~(\ref{RateAllocation}) as
discussed in~\cite{beck2010seq}.

\section{\textcolor{black}{Numerical Results}}
\label{Evaluation}

In this section Monte Carlo simulations are carried out in order to evaluate the system performance of our proposed algorithm. \textcolor{black}{To solve $\textbf{Algorithm 1}$, we use YALMIP toolbox to model the optimization problem with MOSEK as internal solver~\cite{yalmip2004}}. For simulations, we assume that there are two flows from the MBS to two UEs, while the number of available paths for each flow is four \cite{2011PS}. The MBS selects two paths from four most popular paths\footnote{As studied in~\cite{2011PS}, it suffices for a flow to maintain at least two paths provided that it repeatedly selects new paths at random and replaces if the latter provides higher throughput.}. Each path contains two relays, the total number of SCBSs is 8, and the one-hop distance is varying from 50 to 100 meters. The maximum transmit power of MBS and each SC are $43$ dBm and $30$ dBm, respectively, and the SC antenna gain is $5$ dBi. \textcolor{black}{The number of antennas $N_{b}$ at each BS is set to $8$ and $64$ for small and large antenna arrays, respectively.} The number of antennas $N_k$ at UE is set to 2 and 16, for small and large antenna arrays, respectively. The number of RF chains at BS $\radiochain_b$ and UE $\radiochain_k$ are set to $8$ and $2$, respectively.

For simulations purposes, the general channel model for arbitrary antenna arrays is used. In particular, the estimate channel matrix $\hat{{\bf H}}_{(i,j)} \in \mathbb{C}^{N_{i} \times N_{j}}$ of the channel matrix ${\bf H}_{(i,j)} \in \mathbb{C}^{N_{i} \times N_{j}}$ between the transmitter $i$ and the receiver $j$ can be modeled as \cite{Liu2014, 2014jointSM}
\begin{equation}\label{mmWavechannel}
\textcolor{black}{  \hat{{\bf H}}_{(i,j)}=\sqrt{ N_{i} \times N_j } \boldsymbol \Theta_{(i,j)}^{1/2}\left(\sqrt{1-\tau_{j}^{2}}{\bf W}_{(i,j)}+\tau_{j}\hat{{\bf W}}_{(i,j)}\right) }, \notag
\end{equation}
where  $\mathbf{{\bf W}}_{(i,j)} = \big [\mathbf{{\bf w}}^{1}_{(i,j)}, \cdots, \mathbf{{\bf w}}^{n_j}_{(i,j)}, \cdots, \mathbf{{\bf w}}^{N_j}_{(i,j)} \big ] \in\mathbb{C}^{N_{i}\times N_j}$ is the small-scale fading channel matrix, which is independent and identically distributed (i.i.d.) with zero mean and variance $\frac{1}{N_{i}\times N_j}$ in which $\mathbf{{\bf w}}^{n_j}_{(i,j)} \in\mathbb{C}^{N_{i}\times 1}$ is the small-scale fading channel vector between the transmitter antenna array and the $n_j^{\text{th}}$ antenna of receiver $j$. Here, $\tau_{j}\in[0,1]$ reflects the estimation accuracy for receiver $j$, if $\tau_{j}=0$, then $\hat{{\bf H}}_{(i,j)} = {\bf H}_{(i,j)}$, the perfect channel state information is assumed at the transmitters \cite{2006fastCSI}. $\hat{{\bf W}}_{(i,j)}\in\mathbb{C}^{N_{i}\times N_j}$ is the estimated noise, also modeled as a realization of the circularly symmetric complex Gaussian distribution matrix with zero mean and variance of $\frac{1}{N_{i}\times N_j}$ \cite{Vu_LB, Liu2014}. Moreover, $\boldsymbol{\Theta}_{(i,j)}\in\mathbb{C}^{N_{i}\times N_{i}}$ depicts the antenna spatial correlation matrix that accounts for the path loss and shadow fading, such that $\text{Rank}(\Theta_{(i,j)}) \ll N_i$.

We generate the spatial correlation matrix as $\Theta_{(i,j)} = {PL}_{(i,j)} \check{\Theta}_{(i,j)}$ with $\text{Rank}(\Theta_{(i,j)}) = \radiochain_i$, and the normalized spatial correlation matrix with $\text{Tr}(\check{\Theta}_{(i,j)}) = N_i$ \cite{2014jointSM}. The mmWave path loss ${PL}_{(i,j)}$ is modeled as a distance-based path loss for urban environments at $28$\,GHz with a $1$ GHz system bandwidth~\cite{2014mmWavePL, mW2014}, which may exist as a line-of-sight (LOS), non-LOS (NLOS), or blockage states. We adopt the mmWave channel model used in the  system level simulation in~\cite{2014mmWavePL}, given by
\vspace{-0.5em}
\begin{equation}
{PL}(d) = Pr(d){PL}_{\text{LOS}}(d) + (1 - Pr(d)){PL}_{\text{NLOS}}(d), \notag
\end{equation}
where ${PL}_{\text{LOS}}(d)$ and ${PL}_{\text{NLOS}}(d)$ are the distance-based path loss for LOS and NLOS states at distance $d$, respectively~\cite{2014mmWavePL}. Here, $Pr(d)$ denotes a boolean random variable that is $1$ with some probability. For the general blockage channel model, the LOS probability is defined as $\exp(-0.006d)$, then the NLOS probability is $1 - \exp(-0.006d)$~\cite{2014mmWavePL, mW2014}. For the analog beamforming, the side lobe gain $\Gamma$ is set to $\frac{1}{4}$, and the beamwidths at the transmitter and receiver are set to $\frac{\pi}{4}$ and $\frac{\pi}{3}$ radians, respectively.

We assume that the traffic flow is divided equally into two sub-flows, the arrival rate for each sub-flow is varying from $2$ to $5$ Gbps for small antenna array case.  The maximum delay requirement $\beta$ and the target reliability probability $\epsilon$ are set to be $10~\text{ms}$ and $5\%$, respectively \cite{vu2017ultra}. For the learning algorithm, the Boltzmann temperature (trade-off factor) $\kappa_{f}$ is set to $5$, while the learning rates $\learningratei(t)$, $\learningrateii(t)$, and $\learningrateiii(t)$ are set to $\frac{1}{\left(t+1\right)^{0.51}}$, $\frac{1}{\left(t+1\right)^{0.55}}$, and $\frac{1}{\left(t+1\right)^{0.6}}$, respectively \cite{2013backhaul,2018ultra}. The parameter settings\footnote{A simulation source code can be found in \cite{2018Code}, which consists of a set of simple functions that allows to learn the path/route and allocate the transmit power in our paper.} are summarized in Table \ref{table_simul}.
\begin{table}[t]
\caption{Parameter settings} 
\centering
\begin{tabular}{ |l|l|l| }
\hline 
Path loss model~\cite{2014mmWavePL, mW2014} & Values in dB & Bandwidth (W)\\
\hline 
LOS @ 28 GHz         & $61.4 + 20 \log(d)$     & $1$ GHz\\
NLOS @ 28 GHz         & $72 + 29.2 \log(d)$     & $1$ GHz\\
\hline
\end{tabular}
\label{table_simul}
\end{table}
\textcolor{black}{To that end, we would like to notice that our work contains some main features: $(i)$ NUM \cite{neely2010S,2014Uti_delay}, $(ii)$ dynamic path selection learning \cite{bennis2013self}, and $(iii)$ URLLC-aware rate allocation \cite{vu2017ultra}. We consider the following baselines: $\textbf{Baseline 1}$ employs features $(i)$ and $(ii)$ , whereas $\textbf{Baseline 2}$ applies features $(i)$ and $(iii)$, finally $\textbf{Baseline 3}$ considers only feature $(i)$. We benchmark our work and these baselines to assess the impact of the dynamic path selections and of the URLLC-constrained rate allocation, which has not been  addressed in the literature in the context of mmWave communications.} In addition, \textbf{\textit{Single hop}} scheme considers that the MBS delivers data to UEs over one single hop at long distance in which the probability of LOS communication is low, and then the blockage needs to be taken into account \cite{2014mmWavePL}.
\subsection{Small Antenna Array System}
\textcolor{black}{We first evaluate the network performance under the small antenna array setting, i.e., $N_i=8$, $N_j=2$. In Fig.~\ref{avgDelay}, we report the average one-hop delay\footnote{The average end-to-end delay is defined as the sum of the average one-hop delay of all hops.} versus the mean arrival rates $\bar{\mu}$}. As we increase $\bar{\mu}$, \textbf{\textit{baselines}} \textbf{3} , $\boldsymbol{2}$, and $\boldsymbol{1}$ violate the latency constraints at $\bar{\mu}=3.5,\,4.5$, and $5$ Gbps, respectively. While the average delay of our proposed algorithm is gradually increased with $\bar{\mu}$, but under the warming level, $\beta=10~\text{ms}$. \textcolor{black}{The reason is that the delay requirement is satisfied via the equivalent instantaneous rate by our proposed algorithm as per \eqref{eq:delayconst01} and \eqref{eq:delayconst02}, while the \textbf{\textit{baselines}} \textbf{1} and \textbf{3} use the traditional utility-delay trade-off approach without considering the latency constraint, and the \textbf{\textit{baseline}} 2 considers the random \textbf{PS} mechanism only}. The benefit of applying the learning path algorithm is that selecting the path with high payoff and less congestion, results in small latency. Let us now take a look at $\bar{\mu}=4.5$\,Gbps, the average one-hop delay of \textbf{\textit{baseline}} \textbf{1} with learning outperforms \textbf{\textit{baselines}} \textbf{2} and \textbf{3}, whereas our proposed scheme reduces latency by $50.64\%$, $81.32\%$ and $92.9\%$ as compared to \textbf{\textit{baselines}} \textbf{1}, \textbf{2}, and \textbf{3}, respectively. When $\bar{\mu}=5$\,Gbps, the average delay of all \textbf{\textit{baselines}} increases dramatically, violating the delay requirement of $10$\,ms, while our proposed scheme is robust to the latency requirement.

    \begin{figure}[t]
        \centering
        \includegraphics[width=1.1\columnwidth]{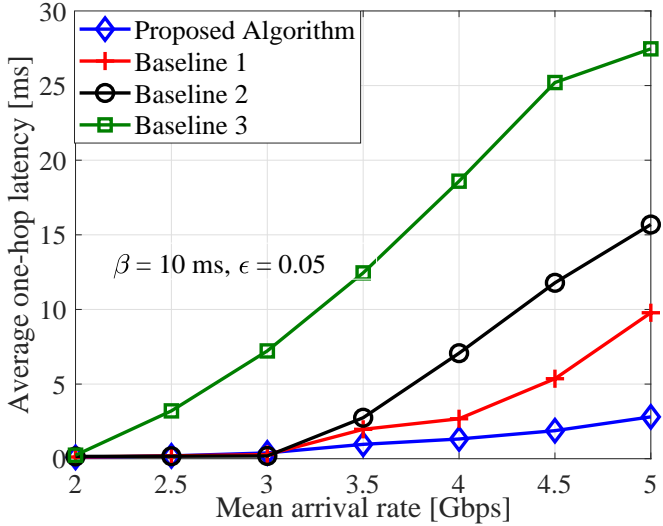}
		\caption{\small Average one-hop delay versus mean arrival rates.}
		\label{avgDelay}
    \end{figure}
	\begin{figure}[t]
        \centering
        \includegraphics[width=1.1\columnwidth]{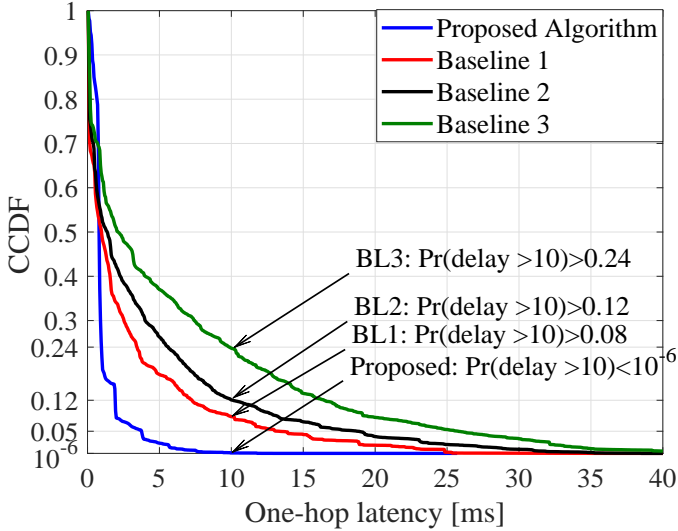}
		\caption{\textcolor{black}{\small CCDF of one-hop latency, small antenna array.}}
		\label{CCDF}
    \end{figure}
    \begin{figure}[t]
		\centering
		\includegraphics[width=1.1\columnwidth]{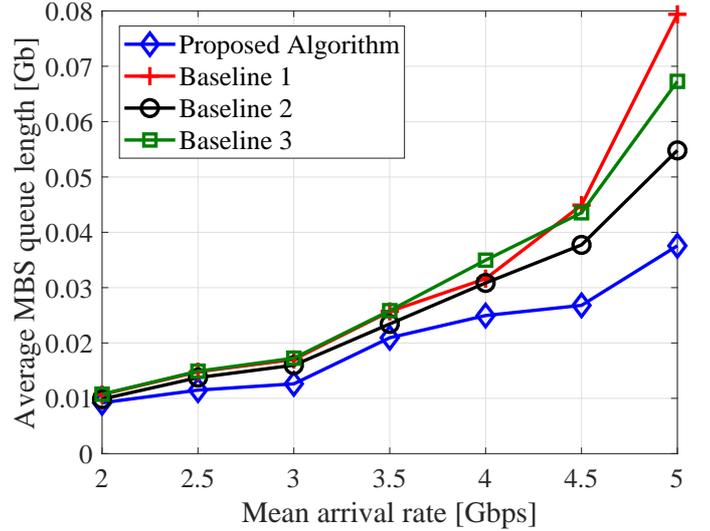}
		\caption{\small Average MBS queue length versus mean arrival rate.}
		\label{avg_MBS_Q}
	\end{figure}

In Fig.~\ref{CCDF}, we report the tail distribution (complementary cumulative distribution function (CCDF)) of latency to showcase how often the system achieves a delay greater than the target delay levels~\cite{2014design} as $\bar{\mu}=4.5$ Gbps, $\epsilon=5\%$, $\beta=10$ ms. In contrast to the average delay, the tail distribution is an important metric to reflect the URLLC characteristic. For instance, at $\bar{\mu}=4.5$\,Gbps, by imposing the probabilistic latency constraint, our proposed approach ensures reliable communication with better guaranteed probability, i.e, $\text{Pr}(\text{delay}>10\text{ms})<10^{-6}$. In contrast, \textbf{\textit{baseline}} \textbf{1} with learning violates the latency constraint with high probability, where $\text{Pr}(\text{delay}>10\text{ms})=0.08$ and $\text{Pr}(\text{delay}>25\text{ms})<10^{-6}$, while the performance of \textbf{\textit{baselines}} \textbf{2} and \textbf{3} gets worse. For instance, as shown in Fig.~\ref{CCDF}, \textbf{\textit{baselines}} \textbf{2} and \textbf{3} obtain $\text{Pr}(\text{delay}>10\text{ms})>0.12$ and $\text{Pr}(\text{delay}>10\text{ms})>0.24$, respectively. For throughput comparison, we observe that for $\bar{\mu}=4.5$\,Gbps, our proposed algorithm is able to deliver $4.4874$ Gbps of average network throughput per each sub-flow, while the \textbf{\textit{baselines}} \textbf{1, 2,} and \textbf{3} deliver $4.4759$, $4.4682$, and $4.3866$ Gbps, respectively. Here, the \textbf{\textit{Single hop}} scheme only delivers $3.55$ Gbps due to the high path loss, causing large latency.

Note that in this work we mainly focus on the low latency scale, i.e., $1-10$ ms, the target achievable rate for all schemes is very high and close to each other. Hence, we report the average MBS queue length instead of the average achievable rate. \textcolor{black}{Generally speaking,} as per \eqref{eq:queue1}, the average achievable rate can be extracted from the average MBS queue length and the mean arrival rate, i.e., $\bar{x}_{f}=\bar{\mu}^{f}-\bar{Q}_{f}$. In Fig \ref{avg_MBS_Q}, we plot the average queue length of the MBS as a function of mean arrival rates. As we increase the mean arrival rate from $2$ to $5$ Gbps, the average MBS queue length of our proposed algorithm is increased from $0.01$ Gb to $0.04$ Gb, which means that the average delay at the MBS is increased from $5$ ms to $8$ ms, which meet the latency constraint \eqref{eq:delayconst}. In contrast, the average queue length of the baselines is increased up to $16$ ms, which violates the latency constraint \eqref{eq:delayconst}.

In Fig.~\ref{CCDF1}, we report the tail distribution of the one-hop latency (in logarithmic scale) versus the guaranteed probability $\epsilon$ as $\beta=10$ ms, $\kappa=5$, and $\bar{\mu}=4.5$ Gbps. By varying $\epsilon$ from $0.05$ to $0.15$, the system is allowed to achieve a delay greater than the target latency with higher probability. As can be seen in Fig.~\ref{CCDF1}, the probability that the system achieves a latency greater than $4$ ms increases from less than $1$ $\%$ to $8$ $\%$ when increasing $\epsilon$ from $0.05$ to $0.15$. This indicates the trade-off between reliability and latency, if we loose the reliability requirement, latency is higher.

\subsection{Large Antenna Array System}
\textcolor{black}{In order to achieve higher beamforming gain, large antenna arrays are employed at both transmitter and receiver, i.e., $N_i=64$, $N_j=16$. In this setting, the maximum transmit power at the MBS is adjusted to $41$ dBm only and the transmitter beamwidth is reduced to $0.5$ radian. Our proposed algorithm is evaluated under both LOS and blockage channel states, whereas all baselines are using the LOS communication model \cite{2014mmWavePL, mW2014, 2018correction}, \cite{2014analysis}. First, in Fig. \ref{CCDF_LS} we plot the the CCDF of one-hop latency (in logarithmic scale) of all schemes when the mean arrival rate is $4.5$ Gbps, which is the same mean admission rate as used in Fig. \ref{CCDF}. Interestingly, due to higher antenna gains all schemes do not violate the latency constraint with an upper bound of $10$ ms and a target probability of $5\%$ as illustrated in Fig. \ref{CCDF_LS}. However, \textbf{\textit{baseline} 3} does not employ the two important features $(ii)$ dynamic path selection learning, and $(iii)$ URLLC-aware rate allocation, and thus, \textbf{\textit{baseline} 3} has a longer tail of latency distribution.}

Next we increase the mean arrival rate to showcase the trade-off between latency and network arrival rate. Fig. \ref{CCDF_L} reports the CCDF of one-hop latency of all schemes with the increasing mean arrival rate, i.e., $\bar{\mu}=9.5$. It can be observed that the performance of our proposed algorithm is degraded under the impact of blockage channels in which the distribution of the latency has a longer tail than \textbf{\textit{baseline}} \textbf{1}. With increasing the mean arrival rate, \textbf{\textit{baselines}} \textbf{2} and \textbf{3} violate the latency constraint with high probabilities, such that $\text{Pr}(\text{delay}>10\text{ms})> 10\%$ for \textbf{\textit{baseline}} \textbf{2} and $\text{Pr}(\text{delay}>10\text{ms})> 20\%$ for \textbf{\textit{baseline}} \textbf{3}. The latency of all schemes increases as we increase the network arrival rate, which states the trade-off between the latency and network arrival rate.

	\begin{figure}[t]\hspace{0.5em} 		
		\centering
		\includegraphics[width=1.1\columnwidth]{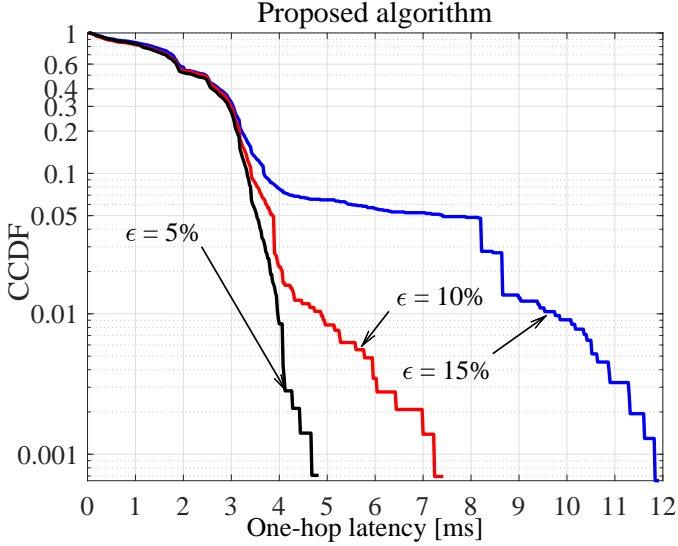}  
		\caption{\small CCDF of one-hop latency versus the guaranteed probability $\epsilon$.} 
		\label{CCDF1}
	\end{figure}

	\begin{figure}[t]
	\centering
    	\includegraphics[width=1.1\columnwidth]{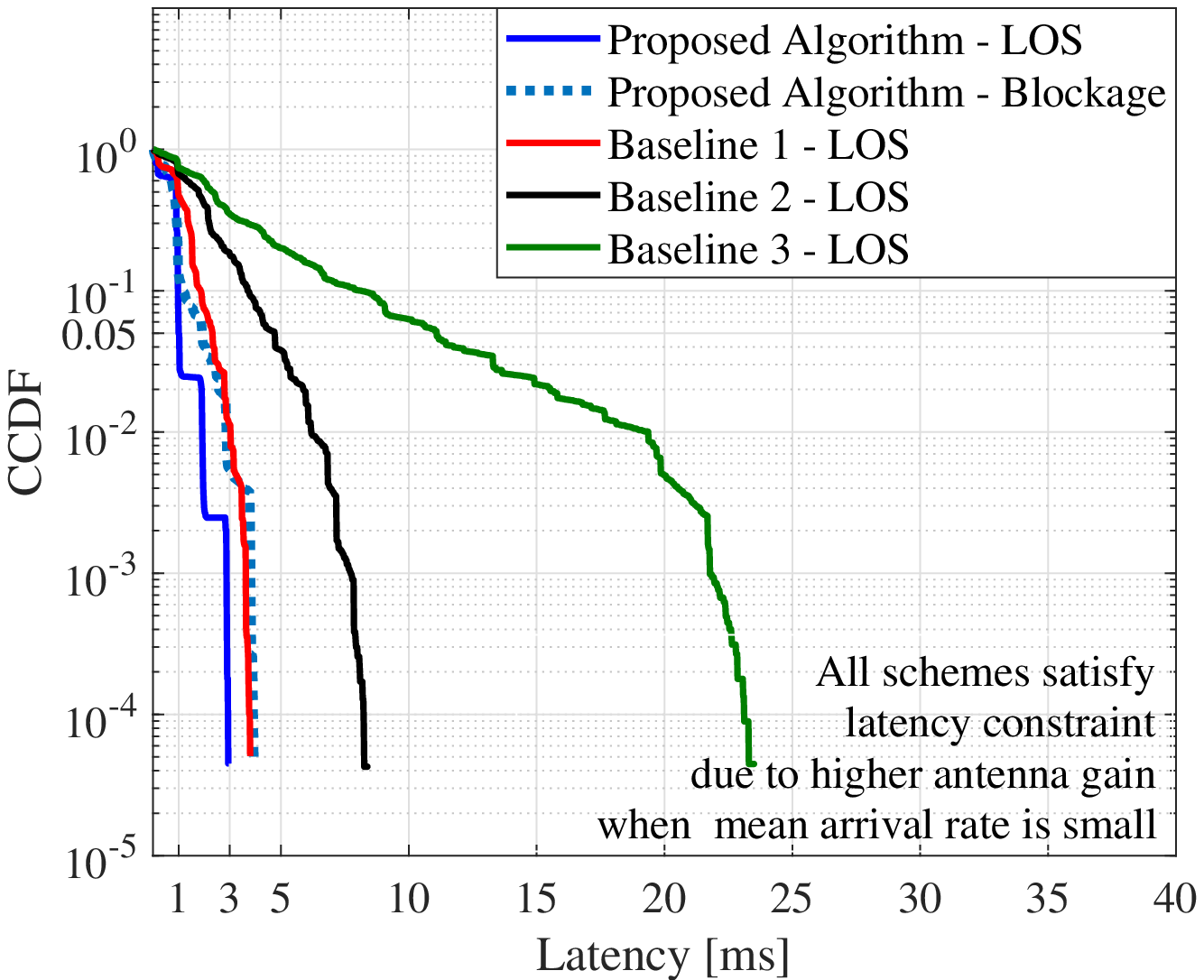}	
    	\caption{\textcolor{black}{\small CCDF of one-hop latency, large antenna array, $\bar{\mu}=4.5$ Gbps.}} 
    	\label{CCDF_LS}
	\end{figure}

	\begin{figure}[t]
    	\centering
    	\includegraphics[width=1.1\columnwidth]{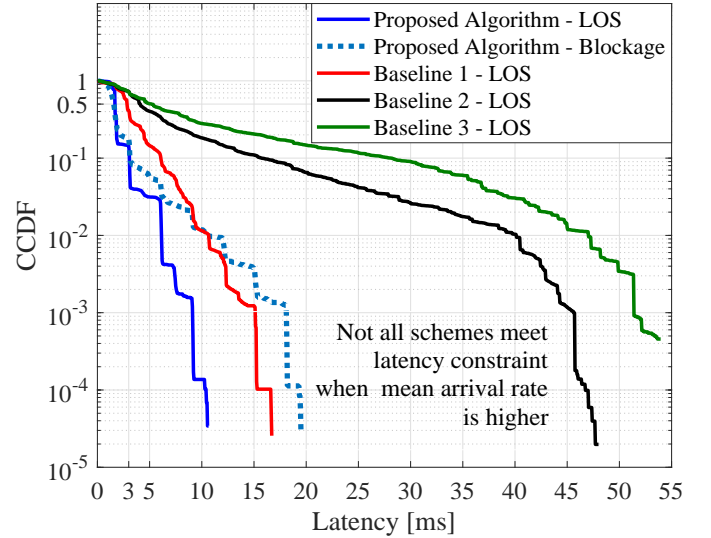}	
    	\caption{\textcolor{black}{\small CCDF of one-hop latency, large antenna array, $\bar{\mu}=9.5$ Gbps.}} 
    	\label{CCDF_L}
	\end{figure}
\subsection{Convergence Characteristics}
We plot the learning convergence of the \textbf{PS} scheme, based on the reinforcement leaning algorithm as shown in Fig. \ref{learning_duration}. In this work, we have applied the Boltzmann-Gibbs technique to capture the trade-off between exploration and exploitation as per \eqref{eq:GibbsDistribution}. We run the simulations for different values of $\kappa\in\{2,5,10,20\}$ for all flow $f$. As expected, with a small value of $\kappa$, the MBS decides to use the paths with highest payoff, selected at the beginning (a small probability of exploration). In this case, the algorithm converges faster, but lacks exploration, the MBS will not try other paths, which may exploit path diversity; as shown in Fig. \ref{avgDelayversusK}, small value of $\kappa$, results in higher delay in the long run. By increasing $\kappa$, the MBS exploits the network environment with higher probability. The benefits of exploration are to utilize the path diversity, improving the performance, i.e., low latency and reducing congestion at the BSs. As shown in Fig. \ref{avgDelayversusK}, average of latency is decreased with $\kappa$, and a large value of $\kappa$ incurs slow convergence.

Next, we plot the convergence of the iterative algorithm as a function of the number of hops as shown in Fig. \ref{learning_duration_iterative}. Here, we provide the distribution of the number of iterations of the SOCP-based algorithm in which the convergence criteria stops running with an accuracy of $10^{-2}$. With increasing the number of hops, the number of constraints and variables is increased, and thus the number of iterations required by the algorithm for convergence is higher. Intuitively, our proposed algorithm only needs few iteration to converge at each time slot $t$ as shown in Fig. \ref{learning_duration_iterative}. For example, for three hop transmission, the probability that the number of iterations takes a value less than or equal to $7$ is $90\%$.

\subsection{Impact of the Learning Temperature}
In addition to the previous discussion on the impact of the trade-off parameter on the convergence, in Fig.~\ref{avgDelayversusK}, we report the average one-hop latency versus the learning trade-off parameter $\kappa$ as $\epsilon=5\%$, $\beta=10$ ms, and $\bar{\mu}=3.5$ Gbps. It can be observed that at small $\kappa$, slowly increasing $\kappa$ the MBS is allowed to explore other paths to get higher gain in the long run. Hence, the average one-hop latency gradually reduces with small increased $\kappa$. However, when $\kappa$ is very large, four paths are determined uniformly for two flows, which becomes random \textbf{PS}. For instance, when $\kappa=50$, the average delay is much higher. Hence, it can be observed that the average delay is a convex function of $\kappa$ in which there exists an optimal value for $\kappa$.

	\begin{figure}[t]
	\centering
	\includegraphics[width=1.1\columnwidth]{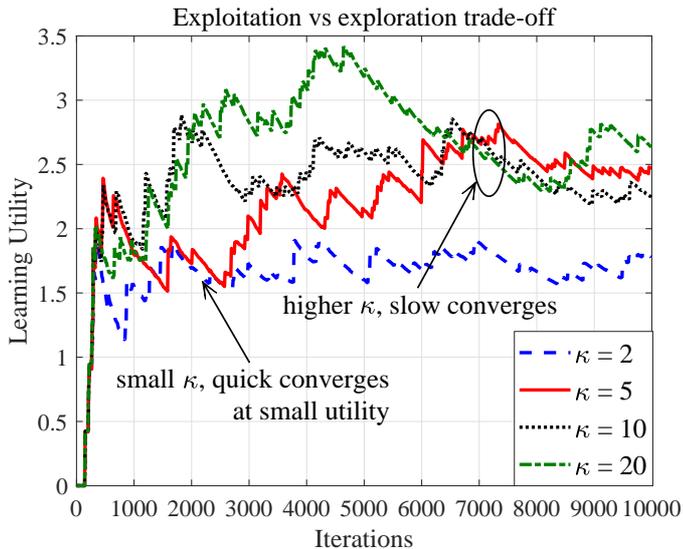}	
	\caption{\textcolor{black}{ \small Convergence of the proposed learning algorithm.}}
	\label{learning_duration}
	\end{figure}
	\begin{figure}[t]
	\centering
	\includegraphics[width=1.1\columnwidth]{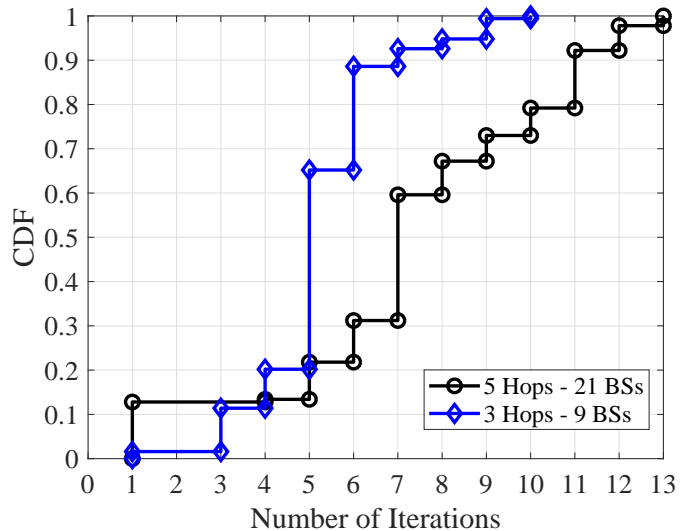}
	\caption{\small The iterative algorithm convergence.}
	\label{learning_duration_iterative}
	\end{figure}
	\begin{figure}[t]\hspace{0.5em} 		
	\centering
    \includegraphics[width=1.1\columnwidth]{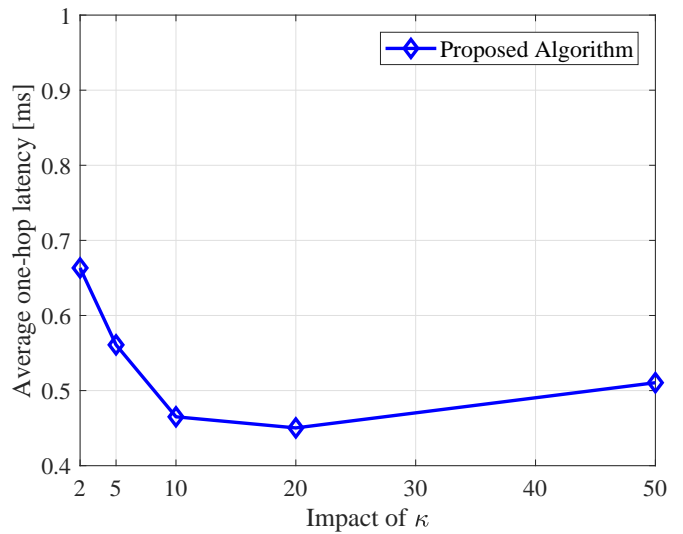}
	\caption{\small Average one-hop delay versus the temperature parameter $\kappa$.}	
	\label{avgDelayversusK}
	\end{figure}

\section{Conclusion}
\label{Conclusion}
In this paper, the authors proposed a multi-hop multi-path scheduling to support reliable communication by incorporating the probabilistic latency constraint and traffic splitting techniques in 5G mmWave networks. In particular, the problem was modeled as a network utility maximization subject to a bounded latency with a guaranteed reliability probability, and network stability. The authors employed massive MIMO and mmWave communication techniques to further improve the DL transmission of a multi-hop self-backhauled small cells. By leveraging stochastic optimization, the problem was decoupled into \textbf{PS} and \textbf{RA}, which are solved by applying the reinforcement learning and successive convex approximation methods, respectively. A comprehensive performance analysis of our proposed algorithm was mathematically provided. Numerical results shown that our proposed framework reduces latency by $50.64\%$ and $92.9\%$ as compared to the \textit{baselines} with and without learning, respectively.

\section*{Acknowledgment}
The authors would like to acknowledge colleagues Chen-Feng Liu, Cristina Perfecto, Mohammed Elbamby, Kien Giang Nguyen, Satya Joshi, Quang-Doanh Vu, Sumudu Samarakoon, and Jihong Park at CWC, University of Oulu for helpful discussions on the paper.

Importantly, the authors also would like to thank the editor and all anonymous reviewers for their constructive comments. They have been very helpful in the revision of this paper and allowed us to improve the technical contents and presentation quality.

\bibliographystyle{IEEEtran}
\bibliography{mmWave}
\newpage  
\begin{IEEEbiography}[]{Trung Kien Vu} 
received the B.Eng. degree from the School of Electronics and Telecommunications, Hanoi University of Science and Technology, Vietnam, in 2012, and the M.Sc. degree in electrical engineering from the School of Electrical Engineering, University of Ulsan, South Korea, in 2014. From 2015 to 2018, he was pursuing the D.Sc. degree with the Centre for Wireless Communications (CWC), University of Oulu, Finland. He was visiting Tsinghua University, Beijing, China from Dec. 2018 to Mar. 2019. His research interests focus on applications of stochastic optimization and deep reinforcement learning for 5G wireless networks and beyond. Kien Vu received the 2016 European Wireless Best Paper Award and was a recipient of the Nokia Foundation grant, the Finnish Foundation grant for Technology Promotion, and the 2014 Brain Korean 21 Plus (BK21+) Scholarship.
\end{IEEEbiography}
\begin{IEEEbiography}[]{Mehdi Bennis}
(S'07--AM'08--SM'15) received the M.Sc. degree from the \'{E}cole Polytechnique F\'{e}d\'{e}rale de Lausanne, Switzerland, and the Eurecom Institute, France, in 2002, and the Ph.D. degree in electrical engineering in 2009 focused on spectrum sharing for future mobile cellular systems. He is currently an Associate Professor with the University of Oulu, Finland, and an Academy of Finland Research Fellow. He has authored or co-authored over 100 research papers in international conferences, journals, book chapters, and patents. His main research interests are in radio resource management, heterogeneous networks, game theory, and machine learning. He has given numerous tutorials at the IEEE flagship conferences. He was also a recipient of several prizes, including the 2015 Fred W. Ellersick Prize from the IEEE Communications Society (ComSoc), the 2016 IEEE COMSOC Best Tutorial Prize, the 2017 EURASIP Best Paper Award for the {\it Journal on Wireless Communications and Networking}, and recently the Best Paper Award at the European Conference on Networks and Communications 2017.
\end{IEEEbiography}
\begin{IEEEbiography}[]{M{\'e}rouane Debbah}
 (S'01--AM'03--M'04--SM'08--F'15) received the M.Sc. and Ph.D. degrees from the {\'E}cole Normale Sup{\'e}rieure Paris-Saclay, France. In 1996, he joined the {\'E}cole Normale Sup{\'e}rieure Paris-Saclay. He was with Motorola Labs, Saclay, France, from 1999 to 2002, and also with the Vienna Research Center for Telecommunications, Vienna, Austria, until 2003. From 2003 to 2007, he was an Assistant Professor with the Mobile Communications Department, Institut Eurecom, Sophia Antipolis, France. From 2007 to 2014, he was the Director of the Alcatel-Lucent Chair on Flexible Radio. Since 2007, he has been a Full Professor with CentraleSup{\'e}lec, Gif-sur-Yvette, France. Since 2014, he has been a Vice-President of the Huawei France Research Center and the Director of the Mathematical and Algorithmic Sciences Lab. He has managed 8 EU projects and more than 24 national and international projects. His research interests lie in fundamental mathematics, algorithms, statistics, information, and communication sciences research. He is an IEEE Fellow, a WWRF Fellow, and a Membre {\'e}m{\'e}rite SEE. He was a recipient of the ERC Grant MORE (Advanced Mathematical Tools for Complex Network Engineering) from 2012 to 2017. He was a recipient of the Mario Boella Award in 2005, the IEEE Glavieux Prize Award in 2011, and the Qualcomm Innovation Prize Award in 2012. He received 19 best paper awards, among which the 2007 IEEE GLOBECOM Best Paper Award, the Wi-Opt 2009 Best Paper Award, the 2010 Newcom++ Best Paper Award, the WUN CogCom Best Paper 2012 and 2013 Award, the 2014 WCNC Best Paper Award, the 2015 ICC Best Paper Award, the 2015 IEEE Communications Society Leonard G. Abraham Prize, the 2015 IEEE Communications Society Fred W. Ellersick Prize, the 2016 IEEE Communications Society Best Tutorial Paper Award, the 2016 European Wireless Best Paper Award, the 2017 Eurasip Best Paper Award, the 2018 IEEE Marconi Prize Paper Award, and the Valuetools 2007, Valuetools 2008, CrownCom 2009, Valuetools 2012, SAM 2014, and 2017 IEEE Sweden VT-COM-IT Joint Chapter best student paper awards. He is an Associate Editor-in-Chief of the journal Random Matrix: Theory and Applications. He was an Associate Area Editor and Senior Area Editor of the IEEE TRANSACTIONS ON SIGNAL PROCESSING from 2011 to 2013 and from 2013 to 2014, respectively.
\end{IEEEbiography}
\begin{IEEEbiography}[]{Matti Latva-aho}
received the M.Sc., Lic.Tech. and Dr. Tech (Hons.) degrees in Electrical Engineering from the University of Oulu, Finland in 1992, 1996 and 1998, respectively. From 1992 to 1993, he was a Research Engineer at Nokia Mobile Phones, Oulu, Finland after which he joined Centre for Wireless Communications (CWC) at the University of Oulu. Prof. Latva-aho was Director of CWC during the years 1998--2006 and Head of Department for Communication Engineering until August 2014. Currently he serves as Academy of Finland Professor in 2017--2022 and is Director for 6Genesis - Finnish Wireless Flagship for 2018--2026. His research interests are related to mobile broadband communication systems and currently his group focuses on 5G and beyond systems research. Prof. Latva-aho has published 350+ conference or journal papers in the field of wireless communications. He received Nokia Foundation Award in 2015 for his achievements in mobile communications research.
\end{IEEEbiography}
\end{document}